\documentclass[reqno]{amsart}
\usepackage{amsmath,amssymb,latexsym}
\usepackage{graphicx}
\usepackage[dvi ps]{epsfig}

\newcommand{\tF}{\tilde F}
\newcommand{\g }{\vert g\vert}
\newcommand{\beq }{\begin{equation}}
\newcommand{\eeq }{\end{equation}}

\def\el{E^{el}}
\def\inel{E^{in}}

\theoremstyle{definition}
\newtheorem{remark}{Remark}
\newtheorem{example}{Example}
\numberwithin{equation}{section}

\begin{document}
\title{A 4D geometrical modeling of a material aging}
\author{ Alexander Chudnovsky}
\address{CEMM, University of Illinois at Chicago, Chicago,
IL}\email{achudnov@uil.edu}
\author{Serge Preston}\address{Department of Mathematics and Statistics, Portland State University,
Portland, OR, U.S.}\email{serge@pdx.edu}

\begin{abstract}
 4-dim intrinsic (material) Riemannian metric $G$ of the material 4-D
space-time continuum $P$ is utilized as the characteristic of the
aging processes developing in the material. Manifested through
variation of basic material characteristics such as density, moduli
of elasticity, yield stress, strength, and toughness., the aging
process is modeled as the evolution of the metric $G$ (most
importantly of its time component $G_{00}$) of the material
space-time $P$ embedded into 4-D Newtonian space-time with Euclidean
metric.\par
 The evolutional equation for metric $G$ is derived by the
classical variational approach. Construction of a Lagrangian for an
aging elastic media and the derivation of a system of coupled
elastostatic and aging equations constitute the central part of the
work. The external and internal balance laws associated with
symmetries of material and physical space-time geometries are
briefly reviewed from a new viewpoint presented in the paper.
Examples of the stress relaxation and creep of a homogeneous rod,
cold drawing, and chemical degradation in a tubing are discussed.
\end{abstract}
 \maketitle

\section{Introduction}

We seek to develop a model of inelastic processes in the aging
materials by employing a 4-D inner material metric tensor $G$ as the
aging (damage) parameter of a material continuum. Aging here implies
any variation in the chemical make-up, i.e., chemical degradation,
phase transformation, phase coarsening, nucleation, and growth of
micro-defects such as dislocations and voids, shear bands, crazes,
micro-cracks, etc. Material engineering and failure analysis
indicate that, in addition to the stress and strain tensors, a
parameter of state (the "aging" parameter) is needed to represent on
a continuum level the sub-micro and micro-structural changes of
material. A kinetic equation for the evolution of the aging
parameter will represent the aging process of a material. The
equations of evolution for the material metric G are the
Euler-Lagrange equations resulting from a Variational Principle. The
conjugate force of the evolution of metric G (and of the related
quantities characterizing the properties of the material) is the
Energy-Momentum Tensor of Elasticity introduced by J. Eshelby (see
Sec.7 below).\par

A 3-D material metric $g$ has long been employed as an internal
variable in continuum mechanics. For example, it was used for
studying the duality of material and physical relations of the
Doyle-Erikson type in article \cite{SM},  the thermodynamics of a
continuum in \cite{CF}, and in \cite{Ep} where the curvature of
material metric $g$ defined by a uniformity mapping of a uniform
material was employed as the driving force of the material
evolution. We use the 4-D material metric $G$ as an additional state
parameter that reflects the aging process. G is introduce with the
largest covariance group allowed by the condition that a small
vicinity of each point of the material preserves its topology during
the aging process (see Sec. 2 below). We consider the 3-D material
metric $g$ on the slices $B_{t}$ of constant physical time as one of
the main dynamical variables following the ADM (Arnovitt, Deser,
Misner) presentation of $G$  (see \cite{MTW,ADM} or Sec.2.3 below).
In that respect, we follow the tradition of the cited works. What is
new in our work is that the smaller (in comparison to the General
Relativity) covariance group of the Lagrangian allows us to use the
lapse function $S$ and the shift vector $\vec N$ as independent
dynamical variables reflecting the proper material time scale and
the intrinsic material flows respectively.\par

This aging parameter is justified by the observation of shrinkage
associated with aging and the subsequent material density variation
as well as a  change of the resonance atomic frequencies and
characteristic relaxation times measured in macroscopical studies.
In other words, the internal length and time scales change with
aging when compared to the corresponding absolute (physical) scales.
The most sensitive indicator of aging is a variation of an intrinsic
material time scale. The measurement of time in the laboratory as
well as in material (intrinsic time) can be accomplished by several
methods, the most common of which is the use of oscillating
processes such as those found in clocks with a pendulum or
crystal-based timepieces. Another way of measuring time is the use
of a unidirectional evolution of state. In medieval Europe, for
example, time was measured by burning a candle which had numbered
and colored beeswax strips. Still another method is associated with
relaxation processes which require an excitation input to enable a
fading response. Electronic relaxation generators employing the
discharge of a capacitor and the fading luminescence of phosphorus
are both examples of relaxation processes, which are well suited for
measuring intrinsic time scale changes because they reflect atomic
or interatomic events.\par

Consider an external excitation of a material which responds with a
specific change in its state. The decay or fading of the response
constitutes the relaxation process. The decay can be described by an
exponential function (within certain limits) $e^{t/\tau_{0}}$ where
$t$ is time and $\tau_{0}$ is the time constant characterizing the
rate of relaxation. Usually $\tau_{0}$ becomes smaller with an
increase in temperature or decrease in pressure. Phosphorous fades
more slowly at colder temperatures, for example.\par

In section 2 we discuss the kinematics of a media with a variable
Riemannian metric $G$ in a 4-D material space-time $P$, embedded
into 4-D Absolute (Newton's) space-time $M^4$ with the Euclidean
metric $H$. Thus, the 3-D "ground state" metric tensor is introduced
together with the proper time lapse function and the material shift
vector field. We consider mass conservation law in section 3.
Elastic and inelastic strain tensors $E^{el};E^{in}$ are introduced
in section 4 as a measure of deformation and the "unstrained state"
respectively. The Lagrangian describing inelastic and elastic
processes in media is discussed in section 5. A variational
formulation of aging theory and the Euler-Lagrange equations
(equations of elasticity coupled with the aging equations) are
considered in section 6. We present the combined system of
elasticity and aging equations in section 7 and discuss special
cases of the aging equations in section 8. Corresponding to the
material and laboratory symmetries, we consider the space and
material balance laws in section 9. In section 10 the
Energy-Momentum balance Law and the decomposition of the
Energy-Momentum tensor into components, including the Eshelby tensor
and terms related to the aging processes are presented. In the final
section we explore the application of this model to the basic
inelastic processes- unconstrained aging, stress relaxation, and
creep of a homogeneous rod.\par

\section{4D kinematics of media with a variable metric.}

In this section we introduce the basic elements of the kinematics of
a continuum with a variable metric, including material space-time
$P$, 4D material metric $G$, 4D deformations $\phi$, slicing of the
material space-time by the surfaces of constant physical time
$B_{\phi ,t}$, and total, elastic and irreversible strain tensors.

\subsection{Physical and Material Space-Time}

Let us consider the 4-D Euclidean vector space $M={\mathbb R}\times
{\mathbb R}^{3}$ ({\bf physical space-time}) with the standard
Euclidean metric $H$. There exists the volume form $d^{4}v$
corresponding to this metric.\par

We select global coordinates $x^{i},i=1,2,3,$ in the physical space
${\mathbb R }^{3}$ and $x^{0}=t$ on the time axes ${\mathbb R}$. We
have $H=dt^2+h=dt^2+\sum_{i}dx^{i\ 2}.$\par
 Hyperplanes $t=c$ are endowed with the 3D Euclidean metric $h$ induced by $H$.
We extend 3D tensor $h$ to the degenerate (0,2)-tensor $\hat{h}$
 in $M$, taking ${\hat{h}}_{0i}={\hat{h}}_{i0}={\hat h}_{00}=0$.
\par
A solid is considered here, in a conventional way as a 3D manifold
$B$ with the boundary $\partial B$, i.e. a set of "idealized"
material points.  We will use local coordinates $X^{I},I=1,2,3$
which, incidently, may be global coordinates induced by a {\bf
reference configuration} i.e., a diffeomorphic embedding $\phi _{0\
3}: B \longmapsto {\mathbb R}^{3}$ (\cite{MH}).
\par

Cylinder $P={\mathbb R}\times {\bar B}$ (with the coordinates
$(X^{0}=T,X^{I},I=1,2,3)$) is equipped with the 4D Riemannian
metric $G$ (material metric) with the components $G_{IJ}$ relative
to the coordinates $X^{0}=T,\ X^I, I=1,2,3$. Space $(P,G)$ is
further referred to as the {\bf material "space-time"}.
\par

Metric $G$ defines the 4D volume form $dV=\sqrt{\vert G\vert
}d^{4}X$, where $\vert G\vert $ is the determinant of the matrix
$(G_{IJ})$.
\par

An example of such a material metric $G$ can be constructed as
follows. Extend the reference configuration $\phi _{0\ 3}$ to the
diffeomorphic embedding $\phi _{0}: P \longmapsto M,\ \phi
_{0}(T,X^I )=(T,x^i=\phi^{i}_{0\ 3}(X^I ))$.  Let $G_{0}$ be the
metric $\phi ^{*}_{0}(H)$ (here and below we denote by $\phi ^{*}Q$
the pullback of a covariant tensor $Q$ by the differentiable mapping
$\phi $). In the coordinates $(X,T)$ the matrix of the metric
$G_{0}$ is
$\left(\begin{array}{cc} 1& 0\\
0&h_{IJ}
\end{array}\right) $.
Denote by $dV_{0}=\sqrt{\vert G_{0}\vert}d^{4}X=\phi
_{0}^{*}(dv^{4})$ the 4D-volume element defined by the metric
$G_{0}$.\par

Projection $\pi : P \longmapsto B$ along $T$-axes plays the same
role in the construction below as in the relativistic elasticity
theory (\cite{CQ},\cite{KM}). In particular, we require invariance
of Lagrangian theory with respect to the automorphisms of the bundle
$(P, \pi , B)$ (diffeomorphisms of material space-time $P$ onto
itself, projecting to $B$, so that material points retains their
identity during the material evolution) preserving the direction of
the flow of the "intrinsic" time (see below), but not with regard to
the whole group of diffeomorphisms of $P$ as in Gravity Theory.

\subsection{Deformation History }

 The {\bf history of the deformation} of the body $B$ is represented by a {\bf
diffeomorphic embedding} $\phi :P \longmapsto M$ of the material
space-time $P$ into the physical (Newtonian) space-time $M$ (see
Fig.1).
\begin{figure}
\scalebox{.75}{\includegraphics{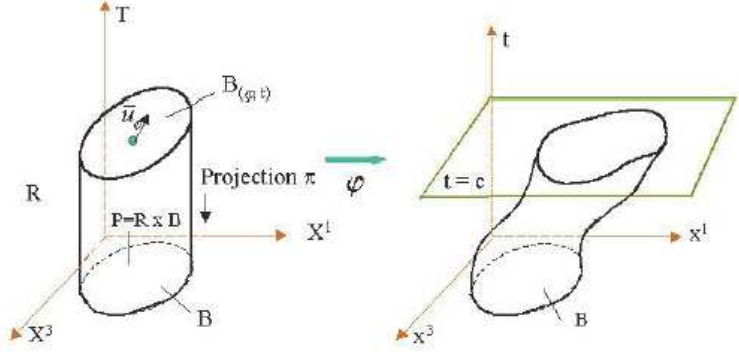}} \caption{4D
Deformation History and the Field of Flow Vector}
\end{figure}

A deformation history $\phi$ for which $t=\phi ^{0}(X)=T$ will be
called {\bf "synchronized"}. The synchronization can practically
performed for relatively slow deformation processes (in comparison
to sound wave velocity).
\par
Using the deformation $\phi $, we introduce the slicing of the
material space-time $P$ by the level surfaces of the zeroth
component of $\phi $
\begin{equation}
B_{\phi }=\{B_{\phi ,t}=\phi ^{0\ -1}(t)\} =\{(T,X)\in P\vert \phi
^{0}(T,X)=t\}.
\end{equation}

 For a synchronized deformation $B_{\phi ,t}=\{(T,X)\in P\vert T=t\}.$ \par

 There is a {\bf time flow vector field} $u_{\phi }$ in $P$,
associated with the slicing $B_{\phi ,t}$ of the space-time $P$
(\cite{MTW}, \cite{CQ}).  This (future directed) vector field
represent the flow of "intrinsic" (proper in Relativity Theory) time
in the material. Lifting the index in the 1-form $d\phi ^{0}$ with
the help of the metric $G$ and normalizing obtained vector field, we
define the time flow vector field as
\begin{equation}
u^{.}_{\phi }=\frac{(d\phi ^{0})^{\#}}{\Vert d\phi ^{0}\Vert
}_{G}.
\end{equation}

The norm of the 1-form $d\phi ^{0}$ is defined as $\Vert d\phi
^{0}\Vert ^2=(G^{AB}\phi ^{0}_{,A}\phi ^{0}_{,B})^{1/2}$ (summation
agreement by repeating indices is used here). Thus, $u_{\phi }$ is
the unit vector $G$-orthogonal to the slices $B_{\phi ,t}$. In the
local coordinates $X^I$,
\begin{equation}
u^{.}_{\phi }=\frac{G^{AB}\phi ^{0}_{,B}}{\sqrt{G^{AB}\phi
^{0}_{,A}\phi ^{0}_{,B}}}\frac{\partial}{\partial X^{A}}.
\end{equation}
\par
 For the synchronized deformations does not depend on $\phi $:
\begin{equation}
u^{.}_{\phi
}=u^{.}_{G}=\frac{G^{I0}}{\sqrt{G^{00}}}\frac{\partial}{\partial
X^{I}}.
\end{equation}
Additionally, if the metric $G$ has the block-diagonal form in the
coordinates $(X^{0}=T,X^I)$ (shortly, BD - metric), we have
$u^{.}_{G}=[G_{00}]^{-\frac{1}{2}}\frac{\partial }{\partial T}.$
\par
 Let $u_{0}=\frac{\partial}{\partial T}$ be the flow vector associated with the
metric $G_{0}$ and the corresponding 3D slicing $B_{0}$.  \par We
require fulfillment of the following condition ensuring the
irreversibility of the flow of time:
\begin{equation}
 <u_{\phi },u_{0}>_{G}\ >0.
 \end{equation}
Deformation history $\phi $ for which the condition (2.5) is
satisfied is called {\bf admissible}.\par
 In coordinates $(X^{I})$ this condition reduces to the following simple inequality
\begin{equation}
\phi ^{0}_{,0} >0 \end{equation}
 and, therefore is a restriction
on the deformation history
 only.\par  For a synchronized deformation history $\phi $, this condition is
trivially satisfied.
\par
Time component $\phi ^{0}$ of the deformation history may be
excluded from the list of dynamical variables by an appropriate
"gauging".  Namely, we use the invariance of Lagrangian under the
automorphisms of the bundle $(P,\pi ,B)$ to make the deformation
history synchronized.\par
  An automorphism $F: P \longmapsto P, X^{I}=F^{I}(Y^{A})$  of
the bundle $(P,\pi ,B)$ determines the diffeomorphism of the base
$B$ that can be considered as a change of variables
$X^{I}=F^{I}(Y^{A})$.
\par

 In the new variables, the condition
(2.6) takes the form $\frac{\partial \phi ^{0}}{\partial Y^{0}}=\phi
^{0}_{,I}F^{I}_{,0}=\phi ^{0}_{,0}F^{0}_{,0}>0$ (since $F^{I},\
I=1,2,3$ do not depend on $Y^{0}$). Thus, the class of admissible
deformation histories $\phi $ is stable under the action of the
subgroup $Aut^{+}(P)$ of all automorphisms of $P$ with
$F^{0}_{,0}>0$. \par  The group $Aut^{+}(P)$ of automorphisms of the
bundle contains two subgroups.  One is the subgroup $TC$ of the
"time change" proper gauge diffeomorphisms $(X^{0}=T;X^{I},I=1,2,3)
\longmapsto (F(T,;X^{J},J=1,2,3),X^{I},I=1,2,3)$ for an arbitrary
smooth function $F(X^{I})$ with $F_{,0}>0.$ The other one (denoted
$D(B)$) consists of the lifts to the slices $B_{\phi ,t}$ of the
manifold $P$ of the orientation preserving diffeomorphisms of the
base $B$ (group of such transformations of $B$ is denoted
$Diff^{+}(B)$). To lift a diffeomorphism we use (diffeomorphic)
projections $\pi _{\phi ,t}=\pi \vert_{B_{\phi ,t}}:B_{\phi ,t}
\longmapsto B$. If $\phi $ is synchronized, lifted diffeomorphisms
do not depend on $T$.\par Any automorphism of the bundle $\phi \in
Aut^{+}(P)$ generates the time independent diffeomorphism $\phi
_{B}$ of the base $B$, that is element of $Diff^{+}(B).$ Lifting
this element to the element of $D(B)$ we represent the group
$Aut^{+}(P)$ as the semi-direct product of the normal subgroup
$TC(P)$ and the subgroup $D(B)$. Thus, we have proved the first of
two following statements
\begin{enumerate}
 \item Automorphisms group
$Aut^{+}(P)$ is the semidirect product of the subgroup $D(B)\sim
Diff^{+}(B)$ of orientation preserving diffeomorphisms of the base
$B$ and the subgroup $TC$ of proper gauge transformations
$\xi_{F}$ of the fibers
\[
\xi_{F}:((X^{0}=T;X^{I},I=1,2,3) \longmapsto
(F(T,;X^{J},J=1,2,3),X^{I},I=1,2,3))\]
 with $F_{,0}>0.$
\item For any admissible history of deformations $\phi$ one can
choose a transformation $\xi_{F}\in TC$ such that the history of
deformation $\phi \circ \xi_{F}$ is synchronized.
\end{enumerate}
\par

To prove the second statement let $\phi $ be an admissible history
of deformation. Define the element $F \in TC(P)$ as follows:
$F:(T,X^{I},I=1,2,3) \longmapsto (\phi
^{0}(T,X^{J},J=1,2,3),X^{I},I=1,2,3)$. Then, $\phi =\phi _{1}\circ
F$ where $\phi _{1}$ is  another admissible history of deformation
with the same components $\phi ^{i},i=1,2,3$ and the identity
component $\phi _{1}^{0}(T,X^{I},I=1,2,3)=T$.
  Transformation $F$ is admissible since $F^{0}_{,0}=\phi ^{0}_{,0} >0$, thus
$F \in TC(P).$ Apparently, the deformation history $\phi _{1}$ is
synchronized.\par

 Therefore we may restrict our consideration to the synchronized histories of
deformation keeping in mind that the covariance group of the theory
reduces from the group $Aut^{+}(P)$ to the group $D(B)$ of
time-independent orientation preserving diffeomorphisms of base $B$.

\subsection{ADM-decomposition of Material Metric, Lapse and Shift.}

\par
 Slicing $B_{\phi ,t}$, generates the (3,1)-decomposition of the material metric
$G$ employed (for a Lorentz type metric) in General Relativity
(\cite{MTW},\cite{ADM}). Specifically, the sandwich structure of the
part of the manifold $P$ bounded by the surfaces $B_{\phi ,t}$ and
$B_{\phi ,t+\Delta t}$ allows one to introduce the time-dependent
lapse function $N$ and the shift vector field $\vec{N}$ tangent to
the slices $B_{\phi ,t}$  such that the metric is block-diagonalized
in the moving coframe $(dT,dX^{A}+N^{A}dT)$:
\begin{equation}
ds^{2}=G_{IJ}dX^{I}dX^{J}=g_{IJ}(dX^{I}+N^{I}dT)(dX^{J}+N^{J}dT)+S^{2}dT^{2}.
\end{equation}

\begin{figure}
\scalebox{.75}{\includegraphics{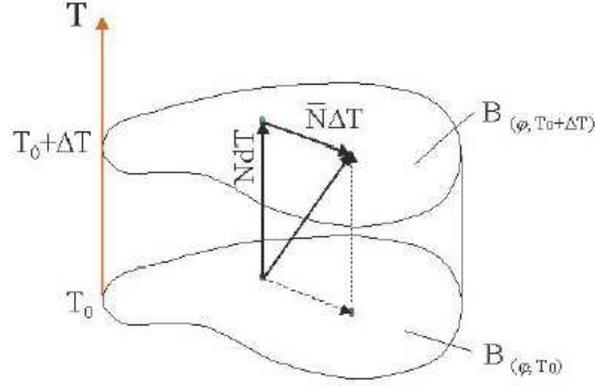}} \caption{ Lapse
Function $S$ and Shift Vector $\vec{N}$. }
\end{figure}

 Matrix representation of the material
metric tensor $G$ and inverse tensor $G^{-1}$ have in these
notations, the forms
\begin{equation}
\left(\begin{array}{cc} G_{00} & G_{0J}\\
G_{I0} & G_{IJ}\end{array}\right) =\left(\begin{array}{cc} N_{A}N^{A}+S^{2} & N_{J}\\
N_{I} & g_{IJ} \end{array}\right) ,\ \ \left(\begin{array}{cc} G^{00} & G^{0J}\\
G^{I0} & G^{IJ}\end{array}\right) =\left(\begin{array}{cc} \frac{1}{S^{2}} & -\frac{N^{J}}{S^{2}}\\
-\frac{N^{I}}{S^{2}} & g^{IJ}+\frac{N^{I}N^{J}}{S^{2}}
\end{array}\right) ,
\end{equation}
 where $g$ is the 3D-metric induced by $G$ on the slices $B_{\phi ,t}$ and
$g^{-1}$ is the corresponding inverse tensor. In these notations $
\sqrt{\vert G\vert }=S\sqrt{\vert g\vert}.$\par In what follows we
assume that the 4D-deformation history $\phi $ is synchronized. Thus
slices $B_{\phi ,t}$ has the form $T=t=const$.
\par In these notations, the flow vector $u_{G}$ and the corresponding 1-form have the form
\begin{equation}
 u_{G}  =  \frac{1}{S}\partial _{T}-\frac{N^{A}}{S}\partial
 _{X^{A}},\ \ u_{G}^{\flat} = SdT.
\end{equation}

The last formula gives the "material time differential" $d\tau =SdT$
for the material metric $G$.  In our context, the coordinate X
dependence of lapse function S(T,X) accounts for heterogeneity of
material aging in different points of the solid. On the Fig. 3 the
local observer at different points of body at different moments of
time $T$ sees the different rate of the local time in comparison
with the laboratory clocks.\par
 Moreover, the lapse function S can be considered as an intrinsic measure
of material age, associated with its cohesiveness. It can be
normalized to be equal 1 in the reference state of the solid. As a
result of energy dissipation in various inelastic processes,
material loses its cohesiveness with aging. In the formalism
presented here it is manifested in slowing down of the material
(intrinsic) time, i.e. increasing of $S(X,T)$.\par
    Here we do not consider thermodynamics. However, monotonic
increase of $S(X,T)$ resembles and can be linked to the principle of
non-negative entropy production of the thermodynamics
  of irreversible processes:
$ \frac{dS_{in}}{dt}>0.$

  \par

The requirement $S>1$ leads to the strong constraints on the form of
the "ground state term" of the material Lagrangian $L_{m}$, (see
Sec. 5).

\begin{figure}
\scalebox{.50}{\includegraphics{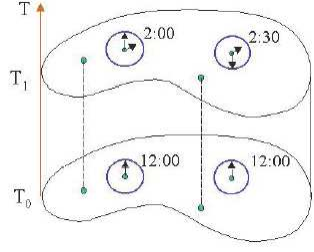}}
\caption{Illustration of Time Rate $S$ (Lapse Function) by Variation
of Time Interval $(\tau -\tau _{0})=S(T_{1}-T_{0}).$}
\end{figure}

 In this context, the shift vector field $\vec{N}$ in the
metric $G$ reflects a propagation of the phase transition or
chemical transformation boundary through the material as reflected,
for instance in the mass conservation law (see below).

\begin{figure}
\scalebox{.50}{\includegraphics{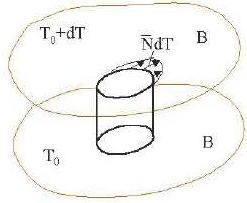}}
\caption{Propagation of Phase Boundary Represented by Shift Vector
$\vec{N}$ }
\end{figure}

The separation of the evolution of 3D material metric $g$, material
transformation process in $B_t$, characterized by $\vec{N}$ and the
inhomogeneity of the local time $d\tau = SdT$, are the main reason
for introduction (3+1) ADT-representation of 4D-metric in Gravity
Theory (\cite{MTW}). In addition, an adoption of this view leads to
a very clear separation of the "physical" degrees of freedom in the
canonical formalism and to the explicit hyperbolic formulations of
Einstein Equations (\cite{ACY,FM}).

\section{Mass conservation law}

 The {\bf mass form} $dM=\rho _{0}dV$ defined in P is  introduced here, in addition
 to the volume form $dV_{G}$ of metric $G$. The reference mass density
$\rho _{0}$, defined by this representation, satisfies the {\bf mass
conservation law} (\cite{MH},\cite{CQ})
\begin{equation}
{\mathcal L}_{u_{\phi }}dM=d(i_{u_{\phi }}dM)=0.
\end{equation}
 Here ${\mathcal L}_{u_{\phi }}$ is the Lie (substantial) derivative of the exterior 4-form
$dM$ in the direction of the vector field $u_{\phi}$. Recall that
the Lie derivative of a differential form $\omega $ along a vector
filed $u$ is defined as ${\mathcal L}_{u}\omega =\frac{d}{dt}\phi
^{*}_{t}\omega \vert _{t=0}, $ where $\phi ^{*}_{t}\omega $ is the
pullback of the form $\omega $ by the flow $t \longrightarrow \phi
_{t}=exp(tu)$ of the vector field $u$ (\cite{MH}).\par

Equation (3.1) is equivalent to the condition $div_{G}(\rho
_{0}u_{\phi })=0$, where divergence is taken with respect to the
volume form $dV$.  In local coordinates the Mass Conservation Law
has the form
\begin{equation} \left(\frac{G^{IB}\phi
^{0}_{,B}}{\Vert d\phi ^{0}\Vert }\rho _{0}\sqrt{\vert G\vert
}\right) _{,I}=0.
\end{equation}
\par
 Due to
the properties of the metric $G$ and the deformation $\phi $, the
material space-time manifold $P$ is foliated by the phase curves
of the flow vector field $u_{\phi }$ and thus the value of the
reference mass density $\rho _{0}(0,X)$ at $T=0$ uniquely defines
its values for all $T>0$.\par

In the synchronized case $\phi ^{0}=T, \Vert d\phi ^{0}\Vert
=\sqrt{G^{00}}=S^{-1},\ G^{M0}=-\frac{N^{M}}{S^{2}},\
G^{00}=S^{-2}$ and (3.2) take the form of the following balance
law
\begin{equation}
(\rho _{0}\sqrt{\vert g\vert })_{,0}=\sum
_{A=1}^{3}\left(N^{A}\rho _{0}\sqrt{\vert g\vert }\right) _{,A}.
\end{equation}
From (3.3) we note that the shift vector field $\vec{N}$ can
describe the {\bf matter (density) flow due to the some internal
processes} such as the phase or chemical transformations.\par
 If, in addition to being synchronized, the material metric $G$ is also in the
BD-form ($\vec{N}=0$), the flow term in (3.3) disappears and the
mass conservation law is equivalent to the following
representation of the reference mass density in terms of its
initial value $\rho _{0}(0,X)=\rho _{0}$:
\begin{equation}
 \rho _{0}(T,X^{I})=\rho _{0}(0,X)\sqrt{\frac{G_{00}}{\vert
G\vert}}=\frac{\rho _{0}(0,X)}{\sqrt{\vert g(T,X)\vert }},
\end{equation}
 where
 $G(0,X)$ is assumed to be Euclidean metric.  If metric $G$ does not changes with
time $T$, we get the classical local mass conservation law
$\frac{\partial \rho _{0}}{\partial T}=0.$\par

\section{Elastic, Inelastic and Total Strain Tensors}

In this section we introduce the principal quantities characterizing
both elastic and inelastic deformation processes. Total deformation
is presented as a composition of elastic and inelastic ones and is
integrable. Its elastic and inelastic "components" are
non-integrable, in general, but might be such in special situations
(see Sec.11). We recall that the presentation of total deformation
as a composition of this type was studied in different forms in many
works (\cite{K,Lee,SO}, to name a few). What is new here is the
4D-approach to this decomposition and direct definition of elastic,
inelastic and total strain tensors in terms of material metric
$g_{t}$ as an independent dynamical variable, reference (undeformed)
Cauchy metric $g_{0}$ and the current Cauchy metric $C_{3}(\phi )$
rather then using the "deformation gradients" (integrable or not) of
elastic and inelastic (plastic) deformations.
\par

 Slicing $B_{\phi \ ,t}$ of $P$ defines
the covariant tensor $\gamma
=G-u_{\phi }\otimes u_{\phi }=^{s}\left(\begin{array}{cc} N_{A}N^{A} & N_{J}\\
N_{I} &g_{IJ} \end{array}\right) $ (\cite{CQ},\cite{MTW}). Here
and later the sign $s$ over $=$ means that this equality is true
in synchronized case.  Tensor $\gamma $ induces the time dependent
3D-metric $g_{t}$ on the slices $B_{\phi \ ,t}$ (see, for example,
\cite{FM}).\par To obtain the expression for $g_{t}$ in material
coordinates $X^{I}$, we notice that the tangent vectors
\begin{equation}
\xi _{I}=-\frac{\phi ^{0}_{,I}}{\phi ^{0}_{,0}}\partial
_{X^{0}}+\partial _{X^{I}},\ I=1,2,3
\end{equation}
form the basis of the tangent spaces to the slices $B_{\phi ,t}.$ In
this basis, $g_{t}$ is given by
\begin{equation}
g_{AB}=g(\xi _{A},\xi _{B})=G_{AB}-G_{0B}\frac{\phi
^{0}_{,A}}{\phi ^{0}_{,0}}-G_{A0}\frac{\phi ^{0}_{,B}}{\phi
^{0}_{,0}}+G_{00}\frac{\phi ^{0}_{,A}\phi ^{0}_{,B}}{\phi ^{0\
2}_{,0}},\ A,B=1,2,3.
\end{equation}

\par
For a synchronized deformation $\gamma = G-(G^{00})^{-1}dT\otimes
dT$ and $g_{t}$ is simply the restriction of 4D-metric $G$ to the
slices $B_{T=c}$, i.e. $g_{t\ IJ}=G_{IJ}\vert_{T=t}$, see (2.8).\par

Denote by $h_{t}$ the 3D metric on the leaves $B_{\phi,t}$ induced
by the metric $G_{0}$ (that is by the tensor $\gamma
_{0}=G_{0}-u_{0}\otimes u_{0}$).\par
Associated with the tensor
$\gamma$ there is the projector ((1,1)-tensor)
\begin{equation}
\Pi=G^{-1}\gamma =I-u^{*}_{\phi} \otimes u_{\phi}=^{s}\ I-\frac{G^{I0}}{G^{00}}\partial _{X^{I}}\otimes dT=\left(\begin{array}{cc} 0 & 0\\
N^{I} & I_{3}
\end{array}\right) .
\end{equation}
on the tangent spaces to the slices $B_{\phi ,t}$, last equality
being true for synchronized deformations $\phi $.\par
 Let us consider the pullback $\ C_{4}(\phi )=\phi ^{*}\hat{h}$ of the degenerate
tensor $\hat{h}$ by the 4D-deformation mapping $\phi $.  Tensor
$C_{4}(\phi )$ is degenerate in $P$, its kernel is generated by
the vector $\phi ^{-1}_{*}(\frac{\partial }{\partial t})$. In
coordinates $(X^{I})$ we have
\begin{equation} C_{4}(\phi )_{IJ}=\left(\begin{array}{cc} h_{ij}\phi
^{i}_{,0}\phi ^{j}_{,0}=\Vert \mathbf{V}\Vert
^{2}_{h} & h_{ij}\phi ^{i}_{,0}\phi ^{j}_{,J}\\
 h_{ij}\phi ^{i}_{,I}\phi ^{j}_{,0} & \sum ^{3}_{i,j=1} h_{ij}\phi
^{i}_{,I}\phi ^{j}_{J}
\end{array}\right) .
\end{equation}
The spatial part of this tensor is the conventional Cauchy-Green
strain tensor $C_{3}(\phi)$ of the Elasticity Theory. Components of
this tensor with indices (0J) and (I0), $I,J=1,2,3$ have the form
{\it velocity\ $\times $ \ deformation covector} (see \cite{E2}).
(00)-component of $C_{4}(\phi )$ is the square of the material
velocity $\mathbf{V}=\phi_{3*}(\frac{\partial}{\partial
T})=^{s}\frac{\partial \phi^{i}}{\partial t}\frac{\partial
}{\partial x^{i}}$.\par \vskip0.5cm
\subsection{Elastic Strain Tensor}

 Here we {\it define} the 4D {\it (1,1)-elastic strain tensor} in $P$.
 We will do it first in linear approximation and then, using logarithm of
 a (1,1)-tensor function, in another way, more
 suitable for large deformations. \par
 We start with the following, conventional definition:
\begin{multline}
{\hat E}^{el\ \cdot }_{\cdot }=\frac{1}{2}G^{-1}(C_{4}(\phi
)-\gamma )=^{s}\\
 =\frac{1}{2} \begin{pmatrix} S^{-2}\Vert
\mathbf{V}\Vert^{2}_{h}-S^{-2}N^{I}\langle
\mathbf{V},\phi^{.}_{,I}\rangle_{h} & S^{-2}\langle
\mathbf{V},\phi^{.}_{,J}\rangle_{h}-S^{-2}N^{K}\langle
\phi^{.}_{,K}\phi^{.}_{,J}\rangle_{h} \\
 - S^{-2} N^{I} \Vert \mathbf{V}\Vert^{2}_{h}-N^{I}+ & -S^{-2}N^{I}\langle
\mathbf{V},\phi^{.}_{,J}\rangle_{h}+g^{IK}C_{3}(\phi)_{KJ}-\\
+(S^{-2}N^{I}N^{K}+g^{IK})\langle \phi^{.}_{,K},\mathbf{V}
\rangle_{h}  & -\delta ^{I}_{J}-S^{-2}N^{I}N^{K}\langle
\phi^{.}_{,K}\phi^{.}_{,J}\rangle_{h}
\end{pmatrix}.
\end{multline}

 This tensor contains the square
of material velocity and the shift vector field. Having in mind
the general, dynamical situation it is more appropriate to use the
following tensor as the proper Elastic Strain Tensor
\begin{equation} E^{el\ .}_{.}(\phi )=\Pi {\hat E}^{el}\Pi = \frac{1}{2}\Pi
G^{-1}(C_{4}(\phi )-\gamma))\Pi =^{s}\frac{1}{2}\left(\begin{array}{cc} 0 & 0\\
g^{IK}\langle \phi ^{.}_{,K},\phi ^{.}_{,0}\rangle_{h}-N^{I}\
&g^{IK}C_{3}(\phi )_{KJ}-\delta ^{I}_{J}
\end{array}\right) ,
\end{equation}
Here $\Pi $ is the projector on the slices $B_{\phi ,t}$ defined in
(4.3). The last equality is valid in the synchronized case. Notice
that the basic invariants $Tr(A^k )$ for the tensor (4.6) are the
same as for the tensor $C_{3}(\phi )-g$.
\par

 For the simplicity we use the same symbol $E^{el}$ for the restriction of
this tensor  to the slices $B_{\phi ,t}$.\par Tensor $E(\phi)^{el}$
is a measure of the deviation of the Cauchy metric $C_{3}(\phi )$ of
the actual state from the "ground state" $G$. For a synchronized
deformation $\phi $ and a material metric $G$ with the zero shift
vector, $E^{el}=\frac{1}{2}(C(\phi )-g_{t})$ has the form of the
conventional elastic strain tensor.\par
\begin{remark}
 The deformation $\phi $ is {\bf essentially 3-dimensional} in
the sense that only the spatial Euclidean metric $g_{0}=h$ in $B$ is
deformed.  The 4D-tensor $C_{4}(\phi )=\phi ^{*}h$ defines the
degenerate metric in the material space-time $P$. It is instructive
to compare $C_{4}(\phi )$ with the (degenerate) tensor $\gamma
=G-u_{\phi }\otimes u_{\phi }$.  The elastic strain tensor $E^{el}$
measures the deviation of $C_{4}(\phi )$ from $\gamma $ on the
slices $B_{\phi ,t}$.  Thus, the scheme presented here is
essentially different from relativistic elasticity theory
(\cite{CQ},\cite{KM}) as well as from 4D version of conventional
elasticity theory.\par

We see from (4.6) that $\el =0$ if and only if the following two
conditions are fulfilled:
\beq
\begin{cases}
1)\ g_{ij}=C_{3}(\phi)_{IJ}=\phi_{3}^{*}(h)_{IJ},\\
2)\ N^{I}=g^{IK}\langle \phi^{.}_{,0},\phi^{.}_{,K}\rangle .
\end{cases}
\eeq

In particular, metric $g$ coincide with the Cauchy metric induced by
deformation $\phi$ and is flat.\par

If $\el=$ then ${\hat E}^{el}=0$ if and only if in addition to the
conditions (4.9) the following conditio is fulfilled \beq 3)\  \Vert
V\Vert^{2}_{h}=g^{IJ}\langle
\phi^{.}_{,0},\phi^{.}_{,I}\rangle\langle
\phi^{.}_{,0},\phi^{.}_{,J}\rangle . \eeq
\end{remark}

 \vskip0.4cm
\subsection{Inelastic Strain Tensor}
Now we introduce the {\it inelastic strain tensor} in linear
approximation
\begin{equation}
{\hat{E}}^{in}=\frac{1}{2}G^{-1}(\gamma -\gamma
_{0})=^{s}\frac{1}{2}\left(\begin{array}{cc} 0 & S^{-2}N^{K}h_{KJ}\\
N^{I} & \delta ^{I}_{J}-g^{IK}h_{KJ}-S^{-2}N^{I}N^{K}h_{KJ}
 \end{array}\right),
 \end{equation}
(last equality being true in synchronized case) and the {\it total
strain tensor} ${\hat{E}}^{tot}$ of the body at each given moment
$T$ to characterize the deviation of the deformed Euclidean metric
$\phi ^{*}h\vert _{B_{\phi ,t}}$ from the initial (Euclidean)
3D-metric $h$ ($h$ being the restriction of $G_{0}$ to the slices
$B_{\phi ,t}$)
\begin{equation}
{\hat{E}}^{tot}=\frac{1}{2} G^{-1}(C_{4}(\phi )-\gamma _{0}).
\end{equation}
 Tensor ${\hat{E}}^{tot}$ can be
represented as the sum of the elastic strain tensor ${\hat{E}}^{el}$
and of inelastic strain tensor ${\hat{E}}^{in}$:
\begin{equation}
{\hat{E}}^{tot}={\hat{E}}^{el}+{\hat{E}}^{in}.
\end{equation}
To obtain the corresponding decomposition for the 3D strain
tensors we apply projector $\Pi$ to the total and inelastic strain
tensors. In particular we introduce
\begin{equation}
E^{in}=\Pi
{\hat{E}}^{in}\Pi=^{s}\frac{1}{2}\left(\begin{array}{cc} 0 & 0 \\
N^{I}-g^{IK}h_{KB}N^{B} & \delta ^{I}_{J}-g^{IK}h_{KJ}
\end{array}\right) ,
\end{equation}
As a result, we get from (4.11) the corresponding decomposition of
"3D total strain tensor" \begin{equation} \Pi E^{tot}\Pi
=E^{el}+E^{in}.
\end{equation}
Restriction of these tensors on the 3D slices $B_{\phi ,t}$ leads
to the more conventional ($t$-dependent) version of this
decomposition.\par For a synchronized deformation history,
restriction of $E^{in}$ to the slices $B_{t}$ takes the form
\begin{equation}
E^{in}\vert
_{B_{t}}=\frac{1}{2}g^{-1}(g-g_{0})=\frac{1}{2}(I-g^{-1}g_{0}),
\end{equation}
that describes the decline of 3D material metric $g$ from its
initial (reference) value $g_{0}=\phi^{*}_{0}h$.\par
 Another way to define strain tensors, more suitable for description of
large deformation is to take
\begin{equation}
 {\hat E}^{el}=\frac{1}{2}ln(G^{-1}C_{4}(\phi ))),
{\hat E}^{in}=\frac{1}{2}ln(G^{-1}_{0}G)), {\hat
E}^{tot}=\frac{1}{2}ln(G^{-1}_{0}C_{4}(\phi ))).
\end{equation}
  We can define
$E^{el},\ E^{in},\ E^{tot}$ correspondingly, using projector $\Pi$.
Strain tensors, defined in such a way will, in some simple cases,
enjoy the same additive relations as (4.11),(4.13).  On the other
case, if elastic deformation happens in the directions different
from the principal axes of inelastic deformation, relation between
these deformations becomes more complex.
\par
 The relationship between these definitions and those of the linear approximation above
 is established by using the fact that for a couple $A,B$ of
(0,2)-tensors such that $A$ is invertible, $ln(A^{-1}B)\approx
A^{-1}(A-B)$ provided $A-B$ is small enough. Thus, when linear
approximation is allowable, first definition is the good
approximation of the second. For instance
\begin{equation}
ln(g^{-1}C_{3}(\phi ))=ln(I+g^{-1}(C_{3}(\phi )-g))\approx
g^{-1}(C_{3}(\phi )-g),
\end{equation}
provided $C_{3}(\phi )-g$ is small.\par \vskip0.5cm
\subsection{Strain Rate Tensor}
 One can also define the
material elastic strain rate tensor as follows
\begin{equation}
{\dot{\hat E}}^{el}={\mathcal L}_{u_{\phi}}{\hat E}^{el},
\end{equation}
as well as inelastic strain rate tensor
\begin{equation}
{\dot{\hat E}}^{in}={\mathcal L}_{u_{\phi}}{\hat E}^{in}.
\end{equation}
 In the case where $G=G_{0}$ and $\phi ^{0}=T$,
elastic strain rate tensor defined in (4.17) has, the same spatial
components as the conventional strain rate tensor (\cite{MH}).\par
Denote by $\dot{G}$ the the Lie derivative ${\dot{G}}={\mathcal
L}_{u_{\phi }}G$ of the metric tensor $G$ with respect to the flow
vector $\vec{u}_{\phi }$.  Then the calculation of the Lie
derivative in (4.17) results in the following relation
\begin{equation} {\dot{\hat E}}^{el}={\mathcal
L}_{u_{\phi}}{\hat E}^{el}=-G^{-1}{\dot{G}}{\hat E}^{el}
+\frac{1}{2}G^{-1}({\dot{C_{4}(\phi )}}-{K}),
\end{equation}
where $K={\mathcal L}_{u_{\phi }}\gamma$ is the {\bf extrinsic
curvature tensor} of the slices $B_{\phi ,t}$.

\begin{remark} Here we are using material coordinates and
tensors only.  In order to obtain the corresponding "laboratory"
quantities (seen by an external observer), one defines the
laboratory (Euler) Elastic Strain Tensor \begin{equation} \epsilon
^{el\ i}_{j}=\phi ^{i}_{,A}\phi ^{-1\ B}_{j}E^{el\ A}_{B}
\end{equation}
 and recalculate all the other quantities accordingly.
\end{remark}

Figure 5 presents the above decomposition of total deformation into
the inelastic and elastic deformations.

\begin{figure}
\scalebox{.50}{\includegraphics{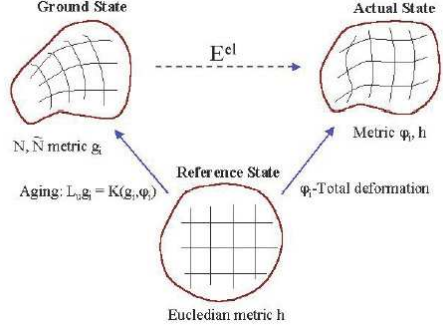}} \caption{
Decomposition of Deformation History.}
\end{figure}

The actual state under the load at any given moment $T$ results from
both elastic (with the variable elastic moduli) and inelastic
(irreversible) deformations. The "ground state" of the body is
characterized by the 3D-metric $g_{t}$.  This state is the
background to which the elastic deformation is added to reach the
actual state (\cite{SO}).\par Transition from the reference state to
the "ground state" that manifests in the evolution of the (initial)
Euclidean metric $h$ to the metric $g_{t}$ cannot be described, in
general, by any point transformation. Transition from the "ground
state" to the actual state at the moment $t$ also is not compatible
in general.  Yet the transition from the reference state to the
actual state is represented by a diffeomorphism $\phi _{t}$.\par

Here we are considering the {\bf material 4D-metric} $G$ and the
{\bf deformation} $\phi $ (or elastic strain tensor $E^{el}(\phi )$)
to be the dynamical variables of the field theory.  The reference
mass density $\rho _{0}$ is found (for the synchronized deformation
$\phi $ and the BD-metric $G$) by the formula (3.4) if its initial
value $\rho _{0}(T=0)$ is known. In this study we consider mainly
the quasi-static version of the theory, i.e. inertia forces and
kinetic energy are assumed to be negligible.

\section{Parameters of Material Evolution, metric Lagrangian}

In examining the processes of deformation and aging of a solid with
a synchronized deformation history $\phi $ we use both general
$(G_{IJ})$ and ADM ($S,\vec{N},g$) notations for the 4D material
metric $G$.
\par
 Following the framework of Classical Field Theory (\cite{BFS}) we take
a  Lagrangian density  ${\mathcal L}(G,\phi )$ referred to the {\bf
volume form} $dV$ as a function of 4D-material metric $G$, its
invariants (with respect to the group of $Diff^{+}(B)$ of the
orientation preserving diffeomorphisms of the base $B$, see above)
and the elastic strain tensor: $ L(G,E^{el})$.

The Lagrangian $L(G,E^{el})dV$ is represented as a sum of the two
parts: the metric part  $L_{m}(G)$ and, as a perturbation of the
ground state, the elastic part $L_{e}(G,E^{el})$ associated with
elastic deformation
\begin{equation}
L=L_{m}(G)+L_{e}(E^{el},G).
\end{equation}

 Metric Lagrangian $L_{m}(G)$ in (5.1) is introduced to account
for the "cohesive energy" or strength of the solid state, the strain
energy of "residual strain" and the energy of the change associated
with a evolution of material properties in time, for instance
material aging processes of phase transitions.
\par
Metric  Lagrangian $L_{m}$  is the sum of several terms with the
coefficients that may depend on the 3D volume factor $\vert
g_{t}\vert $ (actually $\vert g_{t}\vert /\vert g_{0} \vert $) and
the lapse function $S$.  These volume factors are associated with
the solid state ability to retain its intrinsic topology in contrast
to the fluid and gaseous states.\par

First term of $L_{m}$ is the "ground state energy" $F(E^{in},S,\Vert
\vec N\Vert^{2}_{g})$ (shortly GS) - initial ("cohesive") energy
(per unit volume).  \par

   The second (kinetic) term in $L_{m}$ (see (5.4) below) is the function of invariants of the tensor
$K={\mathcal L}_{u_{G}}\gamma $ of extrinsic curvature of the slices
$B_{T}$ in the material space-time $P$ ( \cite{MTW,FM}).\par

In the ADM notations, the (1,1)-tensor $K$ has the following form

\begin{equation}
( K^{I}_{J})=\left(\begin{array}{cc} 0 & 0\\
(*) & S^{-1}g^{IS}(\xi \cdot g)_{SJ}
\end{array}\right) ,
\end{equation}
where $(*)$ represents the terms which do not enter the invariants
of $K$ and $\xi \cdot g= \frac{\partial }{\partial T}g-{\mathcal
L}_{\vec{N}}g$. Lie derivative ${\mathcal L}_{\vec{N}}g$ of the
tensor $g$ with respect to the vector field $\vec{N}$ is calculated
on each 3D slice $B_{T}$ for a fixed $T$.
\par
 In the case of a
block-diagonal metric $G$ (no shift: ${\bar N} =0$), $\gamma
=\left(\begin{array}{cc} 0&0\\ 0& g_{t}\end{array}\right) $;
therefore, $K$ is, essentially, the time derivative of the metric
$g_{t}$:
\begin{equation} K^{I}_{J}=\left(\begin{array}{cc} 0 & 0\\
0 & S^{-1}g^{IA}\frac{\partial }{\partial
T}g_{AJ}\end{array}\right) .\end{equation}
 Therefore, tensor
$K$ represents the rate of change of intrinsic length scales that
reflects the aging processes. It shall be noted that $K$ is also
related to the elastic strain rate (see (4.19)).\par

$L_{m}(G)$ also may depend on the shift vector field $\vec{N}$
through its norm $\Vert \vec N \Vert_{g}^2=N_{A}N^{A}$ (entering the
"ground state energy" $F$) and, possibly, divergence
$div_{g}({\vec{N}})$ and the "proper time derivative" ${\mathcal
L}_{u}{\vec{N}}$.\par
 We also include the term reflecting the residual strain energy
(incompatibility) of $E^{tot}$ which is accounted for by the
scalar curvature $R(g)$ of 3D material metric $g$.\par

 Summarizing the above assumptions we construct the metric Lagrangian
$L_{m}$ as  the scalar function of the parameters listed above:
\begin{equation}
L_{m}=F(S,\Vert \vec N \Vert^2_{g} ,\inel )+\chi(K)+\alpha
div_{g}(\vec N)^2+ \beta R(g_{t}).
\end{equation}

  Function $\chi$ of invariants of tensor $K^{I}_{J}$ ("dissipative potential",
 comp.\cite{M2}) corresponds to the energy of inelastic processes in the material
\renewcommand{\thefootnote}{\ensuremath{\alph{footnote}}}
\footnote{ Initially (\cite{CP2}) we've considered function $\chi $
to be a quadratic function of invariants $Tr(K),\ Tr({K^2})$ but as
the examples of stress relaxation and creep in a rod demonstrate
this function should be chosen differently, corresponding to the
material studied.  In particular, if the Dorn relation ${\dot
\eta}=D(exp^{\beta \sigma}-1)$ between the stress $\sigma $ and the
strain rate $\dot \eta$ ($\eta (t)$ is the volume preserving part of
inelastic strain ) is to be obtained, one should take $\chi (x) =
cx+\frac{x}{\beta
D}ln(\frac{x}{D})-\frac{1}{\beta}(1+\frac{x}{D})ln(1+\frac{x}{D})$.
}.

Coefficients  $\alpha ,\beta $ may depend on
 $S$ and $\vert g\vert $.
\par
In the case of a homogeneous media or in 1D case, the scalar
curvature $R(g_{t})$ of the metric $g_{t}$ is zero and the
corresponding term in (5.4) vanishes.\par

Given the diversity of the material properties and the different
conditions (of loading, boundary, forces, heat,etc.) of inelastic
processes affecting the material it is especially important to
choose the material Lagrangian of different materials appropriately.
It appears as if the different conditions activate different
 "layers" of structural changes for a given material
and, correspondingly, turn on terms in the "ground energy" and in
the "dissipative potential" that are responsible for the given type
of aging. For example, the slow process of unconstrained aging in a
homogeneous rod  (see sec.11. or \cite{CP3}) is overcome by the
scale processes of a stress relaxation or creep each of which begins
in a different loading situation after the strain energy (density)
reaches a (different) activation level. For these two processes both
ground energy $F$ and the dissipative potential $\chi (K)$ have the
same form different from those for unconstrained aging.\par

\begin{remark}
The scalar curvature $R(G)$ of the 4D metric $G$ can be expressed,
by the Gauss equation, as the combination of scalar curvature of 3D
metric $g$ and of invariants of its {\bf extrinsic curvature}:
$R(G)= -(tr({K}^{2})-(tr K)^{2})+R(g_{t})$, up to a divergence term
(\cite{FM}). As a result, the above form of Lagrangian for an aging
media (5.4) is a generalization of the Hilbert-Einstein Lagrangian
$R(G)\sqrt{\vert G\vert }$ of the General Relativity (\cite{MTW}).
By breaking of the invariance group of general relativity to the
smaller group of automorphisms of the bundle $P\rightarrow B$ we can
use more general form of metric Lagrangian.
\end{remark}

\par
 The perturbation of Lagrangian due to elastic deformation
is taken in the form of the Lagrangian of Classical elasticity
(\cite{MH}, Sec.5.4)
\begin{equation}
L_{e}(E^{el},G)=\frac{\rho _{0}}{2}\Vert V\Vert ^{2}_{h}-\rho
_{0}f(E^{el},G)-\rho _{0}U\circ \phi , \end{equation}
 where $\rho
_{0}\Vert V\Vert ^{2}_{h}=\rho_{0} \sum _{ij} h_{ij}\phi
^{i}_{,0}\phi ^{j}_{,0} $ is the density of kinetic energy, $f$ is
the strain energy per unit of mass,  $U$ is the potential of the
body forces. Strain energy $f$ is assumed to be a function of two
first invariants of the (1,1)-strain tensor $E^{el}$. Strain energy
may depend on the metric $G$ through the invariants of
$g_{0}^{-1}g_{t} ,\ S$, vector field $\vec N$, scalar curvature
$R(g)$ etc.
\par
Because we are considering a quasistatic synchronized theory here
 we ignore the inertia effects and, therefore, omit
the kinetic energy term in (5.5).\par

The strain energy density in linear elasticity is conventionally
presented as follows
\[ f(E^{el})=\frac{\mu }{2}Tr(E^{el\
2})+\frac{\lambda }{2}(Tr(E^{el}))^{2},\] where $\mu, \lambda $ are
the initial values of the Lame constants (\cite{LL}).\par

We assume that Strain Energy $f$ and the "ground state" term $F$ are
independent of each other. Yet, in Appendix A we introduce a scheme
where elastic deformation (elastic strain tensor $E^{el}$) is
considered as (small) perturbation of (large) inelastic deformation
(presented by inelastic strain tensor $E^{in}$).  Therefore, strain
energy $f(E^{el})$ is obtained by decomposition of the function
$F(S,E^{tot})$ into the "Taylor series" by the parameter $E^{el}$.
This leads to the expression of elastic moduli of a media through
the invariants of material metric $g$ and the lapse function
$S$.\par

\section{Action, boundary term, Hooke's law.}

 The Action
functional is the integral of the Lagrangian density $\mathcal
L(G,\phi )$ over a 4D domain $U=[0,T_{0}]\times V$. Here
$(V,\partial V)$ is an arbitrary subdomain of $B$ with the
boundary $\partial V$, combined with the 3D boundary integral that
accounts for the work $W$ of surface traction (\cite{MH}).

\begin{equation}
A_{U}(G,\phi )=\int _{U}(L_{m}(G)+L_{e}(E^{el},G))dV+\int
_{[0,T_{0}]\times \partial V_1}V_{\tau}(\phi ,G)d^{3}\Sigma .
\end{equation}
Here $d^{3}\Sigma $ is the area element on the 3D boundary
$\partial U$ of the cylinder $U$ (\cite{MH}).\par

 The second term on
the right represents a boundary conditions put on the deformation
history $\phi$. Typically the boundary $\partial V$ of the domain
$V\subset B$ is divided into two parts $  V=\partial V_{1}\cup
\partial V_2$.  The deformation is prescribed on the part $\partial V_2$:
$\phi_{\partial V_2}=\psi (t,X), $ while along the part ${\partial
U_1}$ of the boundary the traction $\tau$  is prescribed.  Function
$V_{\tau}$ depends (conventionally) on the velocity $\vec V$ of
deformation $\phi$ and on the traction 1-form $\tau $, which is
chosen in such a way as to have $-\nabla _{\phi }V_{\tau }=\tau$ -
traction (\cite{MH}). In Euclidean space with the dead load one
takes $V_{\tau}=-\tau \cdot \phi $.\par

Deformation $\phi(0,X)$ and the velocity $\vec V=
\phi_{*}(\frac{\partial}{\partial t}$ of material points is
assumed to be given at the moment $t=0$. This determines initial
conditions for the deformation history.\par

The boundary conditions for the metric $G$ (including initial
conditions for 3D material metric $g$) require some special
attention. Initial values of $S,g,\vec N$ are known - prescribed by
the material manufacturing process and by the previous history of
the material deformation. On the part $\partial V_1$ of lateral
surface $\partial V$ where deformation $\phi \vert_{\partial V_1}$
is prescribed (for instance when this part of surface is not moving
at all, see \cite{CP3} for examples) we can find
$\phi_{*}(g\vert_{\partial V_1})$ by measuring distances between the
material points on the boundary of the body in the physical space at
moment $t$ and recalculating them back to $B$ by the tangent to the
(prescribed) mapping $\phi_{t}$. If a part of surface is free from
load, one can use the natural (Neumann type) condition that the mean
curvature (with respect to the metric induced by $g_{t}$) of this
part of surface is zero.  Along the part $V_2$ of the surface where
the load $\tau$ is applied one may use for $g$ analog of
Laplace-Young condition for liquid surfaces relating difference of
pressure with the surface tension and the mean curvature. The
formulation of corresponding boundary conditions are the subject of
another work.\par

From the requirement that the variation of the action near the
lateral sides of cylinder $U$ are zero we get to the natural
boundary condition
\[
P\cdot N=\tau,\  or\ \sum _{IJ}g_{IJ}P^{I}_{j}N^{J}=\tau _{j}
\]
in terms of the first Piola-Kirchoff stress tensor $P^{I}_{j}$
defined by the equation (material form of the Hooke's law, see
\cite{MH}):
\begin{equation} P^{I}_{j}=-\frac{\partial L_{e}}{\partial \phi ^{j}_{,I}}=\frac{\partial f}{\partial \phi ^{j}_{,I}}.
\end{equation}
Notice that if the kinetic energy is included into $L_{e}$,
Piola-Kirchoff Tensor has the density of linear momentum vector as
its $P^{0}_{i}$ components (\cite{M,E3}).

We will be using the second (material) Piola-Kirchoff tensor
$S^{I}_{J}=P^{I}_{i}\phi^{i}_{,J}$.\par
 It is useful to recall that the (laboratory) Cauchy stress tensor
$\sigma _{ij}$ is related to the first Piola-Kirchoff tensor by the
following formula $\sigma_{ij}=J^{-1}(\phi)h_{ik}\phi
_{,I}^{k}P^{I}_{j}, $ $J(\phi)$ being the Jacobian of the
deformation $\phi $.\par The zero condition for the variation at the
top ($T=T_{0}$) and the bottom ($T=0$) of the cylinder lead to the
relation between the linear momentum and the kinetic energy in the
classical case (Legendre transformation).  In the scheme presented
here these variations also includes terms related to the aging
processes.\par

\section{Euler-Lagrange Equations.}
The variation principle of the extreme action $\delta A=0$ taken
with respect to the dynamic variables $\phi $ and $G$ results in a
system of Euler-Lagrange equations that represent the coupled
Elasticity and "Aging" equations

\begin{eqnarray}
 \frac{\partial }{\partial T}(\rho _{0}\phi
^{m}_{,0})+ \frac{\partial {\mathcal L}_{e}}{\partial \phi
^{m}}-\sum _{I=1}^{I=3}\frac{\partial }{\partial X^{I}}\left(
\frac{\partial {\mathcal L}_{e}}{\partial \phi ^{m}_{,I}}\right)
-\rho _{0}\sqrt{\vert G\vert }(\nabla B)_{m}  =  0,\
m=1,2,3.  \\
\frac{\delta {\mathcal L}_{m}}{\delta G_{IJ}}=- \frac{\delta
{\mathcal L}_{e                            }}{\delta
G_{IJ}}=\sqrt{\vert G\vert }T^{IJ},\   I,J=0,1,2,3.
\end{eqnarray}

{\bf The Elasticity Equations} (7.1) are obtained by taking the
variation $\delta A$ with respect to the components $\phi ^{i}$
within the domain $U$.  In the case of a BD metric $G$ and the
synchronized deformation $\phi $, these equations coincide with the
conventional dynamical equations of Elasticity Theory. However their
special features are associated with the different form of the
elastic strain tensor $E^{el}$ and with the dependence of the
elastic parameters on time through the invariants of the metric $G$.
The evolution of these parameters is defined by the equations (7.2)
(referred to as Aging equations).\par

{\bf The Aging Equations} (7.2) resulting from the variation of
action with respect to the metric tensor $G$ describe the evolution
of the material metric $G$ for a given initial and boundary
conditions.\par

 The right side of the equations (7.2) represents the
(symmetrical) {\bf "Canonical Energy-Momentum Tensor"} $\sqrt{\vert
G\vert} {T}^{IJ}=-\frac{\delta {\mathcal L}_{e}}{\delta G_{IJ}}$
(\cite{ADM}). In our situation this tensor is closely related to the
{\bf Eshelby EM Tensor} $b_{IJ}$.\par  In his celebrated works
J.Eshelby (\cite{E1,E2}), introduced the 3D and then 4D dynamical
energy-momentum tensor ({\bf Eshelby EM Tensor}) $b$ (denoted
$P^{*}_{lj}$ in \cite{E2}).
\begin{equation}
 b^{I}_{J}=f\delta ^{I}_{J}-\sum _{i=1}^{i=3}\frac{\partial f}
 {\partial \phi ^{i}_{,I}}\phi ^{i}_{,J}=f\delta ^{I}_{J}-S^{I}_{J},
\end{equation}
$f$ being the elastic energy per unit volume.\par
  The tensor $b$ includes the 3D-Eshelby stress tensor (\cite{E2,E3})
, the 1-form of {\it quasi-momentum (pseudomomentum)} ${\mathcal
P}=b^{0}_{J},\ J=1,2,3$, (see \cite{E2,M}), strain energy density
$b^{0}_{0}=-{ L}_{e}=f$ (plus kinetic energy, if the last one is
present) and the energy flow vector $s=b^{I}_{0}=-P^{I}_{i}\phi
^{i}_{,0}, \ I=1,2,3$.  In the quasi-static case $b^{0}_{J}=0$ for
$J=1,2,3.$  In the case of a BD metric (${\vec{N}}=0$) $G$ we have
${\mathcal P}_{B}=b_{0B}=0,\ B=1,2,3$.\par
  Tensor $b_{IJ}$ is, in general, not symmetric (although its 3x3
space part is symmetric with respect to the Cauchy metric
$C_{3}(\phi )$, see \cite{M}.
\par
It was proved in \cite{CP1} that if metric $G$ is block diagonal
(i.e. if $\vec N=0$) and the body forces are zero, then

\beq
 T^{IJ}=\frac{1}{2}b^{(IJ)}+\left( \frac{\delta f(S,g,E^{el})}{\delta g_{IJ}}\right) _{exp},\ I,J=1,2,3,
\eeq

 where $b^{(IJ)}$ is the symmetrical part of the 4D Eshelby
tensor and the symbol $_{exp}$ refers to the derivative of $L_{e}$
by the {\bf explicit dependence of $G$} (not through $\el$).\par

For the Lagrangian $L=L_{m}+L_{e}$ defined by (5.4-5) the Aging
Equations (7.2) can be rewritten in the more convenient ADM
notations.

 The aging equations (7.2) take the form
of the system of PDE for the lapse function $S$, shift vector field
$\vec{N}$ and the 3D material metric $g$. The explicit form of the
above equations can be readily obtained for the Lagrangian in a form
(5.4-5).  In order to achieve this the variational derivatives of
components of Lagrangian with respect to the variables $S,\vec N
,g_{IJ}$ need to be calculated. In Appendix B we calculate the
variations of some of these terms and present them in tabular form.
\par

Variation by $S$ (assuming that $f$ does not depend on $S$):
\begin{equation} \frac{\delta{ \mathcal L}}{\delta S} =0\Leftrightarrow
(F+S\frac{\partial F}{\partial S})+(\chi(K)-\frac{\partial
\chi}{\partial K}:K)+\alpha div_{g}({\vec N})^2+\beta R(g)=
f+S\frac{\partial f}{\partial S}.
\end{equation}

Variation by $N^{I}$:
\begin{multline}
\frac{\delta{ \mathcal L}_{g}}{\delta N^{I}} =2\frac{\partial
F}{\partial (\Vert {\vec N}\Vert ^{2})}N_{I}-\frac{\partial
}{\partial X^{I}}(2\alpha \cdot ln(\rho_{0}S)\cdot div_{g}({\vec
N}))+[ -S^{-1}\frac{\partial \chi }{\partial
K^{A}_{J}}g^{AS}\partial_{X^{I}}g_{SJ}+
 \\
 +\frac{1}{\rho_{0}S\sqrt{\vert g\vert }} \partial_{X^{S}}(\rho_{0}\sqrt{\vert g\vert }
 \frac{\partial \chi }{\partial K^{A}_{J}}g^{AS}g_{IJ})+ \frac{1}{\rho_{0}S\sqrt{\vert
g\vert }}\partial_{X^{J}}(\rho_{0}\sqrt{\vert g\vert }
\frac{\partial \chi }{\partial K^{I}_{J}})] =0.
\end{multline}

Variation by $g_{IJ}$ (for simplicity, we omit in this equation the
terms coming from $div_{g}({\vec N})^2$ in Lagrangian (5.4-5), for
the corresponding term in the equation see Appendix B):

\begin{multline}
[-\beta {\mathcal E}^{AB}+ \frac{1}{\rho_{0}
S}(\Delta_{g}(\rho_{0}\beta S)g^{AB}+
Hess^{AB}(\rho_{0}\beta S)) ] +\\
+[-\frac{\partial \chi }{\partial K^{I}_{J}}
S^{-1}g^{IA}\frac{\partial N^{B}}{\partial X^{J}}-\frac{\partial
\chi }{\partial K^{I}_{B}} S^{-1}g^{IS}\frac{\partial
N^{A}}{\partial X^{S}}- \frac{\partial \chi }{\partial
K^{I}_{J}}g^{IA}K^{B}_{J}- \frac{1}{\rho_{0}S\sqrt{\g
}}\partial_{t}\left( \rho_{0}\sqrt{\g }\frac{\partial \chi
}{\partial
K^{I}_{B}}g^{IA}\right) +\\
+ \frac{1}{\rho_{0}S\sqrt{\g}}\partial_{X^{K}}\left(
\rho_{0}\sqrt{\g} \frac{\partial \chi }{\partial K^{I}_{B}}g^{IA}
N^{K} \right) ]+ \frac{1}{2}L_{m}g^{AB}+\frac{\partial F}{\partial g_{AB}}+ \\
+\frac{\partial F}{\partial \Vert {\vec N}\Vert^{2}_{g}}\cdot
[N^{A}N^{B}+\frac{1}{2} \Vert {\vec N}\Vert^{2}_{g} g^{AB}] =
\frac{1}{2}(f+U)g^{AB}-\frac{1}{2}S^{(AB)}+\frac{\partial
f}{\partial g_{AB}}\text{\Small
exp}=\frac{1}{2}b^{(AB)}+\frac{\partial f}{\partial
g_{AB}}\text{\Small exp}+\frac{1}{2}Ug^{AB}.
\end{multline}

Here ${\mathcal E}^{AB}=Ric(g)^{AB}-\frac{R(g)}{2}g^{AB}$ is the
Einstein tensor of metric $g$.  In the first line of equations (7.6)
(left side) $\Delta _{g}$ is the 3D Laplace operator is defined by
the metric $g$, $Hess(f)=f_{;M;N}$ stands for the Hessian of the
function $f$ (double covariant derivative tensor of $f$).
\par
  On the right side of (7.7) remains the symmetrized Second
Piola-Kirchoff Stress Tensor $S$ (Here and thereof
$S_{(AB)}=\frac{1}{2}(S_{AB}+S_{BA})$) or the Eshelby stress tensor
since  $S^{(AB)}=-b^{(AB)}+{ \mathcal L}_{e}g^{AB}$.  The Eshelby EM
Tensor $b$ is thus the driving force of the evolution of material
metric $g$ (comp. \cite{EM}).\par

 Equations (7.1-2) together with the equation (3.4) for the reference
density form a closed system of equations for dynamic variables
$(G_{IJ},\phi ^{i},\rho _{0}).$  Complemented with the initial and
boundary conditions, these equations provide a closed non-linear
boundary value problem for the deformation of solid and evolution of
the material properties.
\par
In general, the system (7.1-2) seems rather complex, especially if
$L_{e}$ depends on the metric $G$ an its (differential) invariants
explicitly. Nevertheless, leaving a detailed analysis of this system
for future studies, we make some brief remarks about special cases
where system (7.2) is effectively simplified.\par

\section{Special cases and examples.}

\subsection{ Block-diagonal metric $G$}
 In a case of a BD-metric, ${\vec{N}}=0$ (no shift). Therefore, the metric Lagrangian has
the form $L_{m}=F( S,\inel )+\chi (K)+\beta R(g)$ that includes time
derivatives of 3D metric $g$ (in $\chi(K)$) and the space
derivatives of $g$ in the curvature term $R(g_{t}).$\par
 No derivatives of the lapse function $S$ appear anywhere in Lagrangian. In
particular, equation obtained by variation of $S$ {\bf is not a
dynamical equation but rather a constraint}, similar to the
"energy constraint" in the Einstein equations (\cite{FM}).\par
  In the case, when the elastic coefficients do not depend on $S$,
equation (7.5) takes the form
\begin{equation} (F+S\frac{\partial F}{\partial S})+(\chi(K)-\frac{\partial
\chi}{\partial K}:K)+\beta R(g)= f,
\end{equation}
where $\rho _{0}f\sqrt{\vert g\vert }S$ is the density of strain
energy (per unit of unperturbed volume).\par
 This relation represents an equilibrium between the strain energy in the material
 (residual stresses
 presented by the scalar curvature of $g$)
 and the internal material constituents  (the "ground state term" and the terms
 defined by the kinetic of material processes).  In the case of a
 homogeneous tensile rod (\cite{CP3}) this relation determines the domain of
 admissible evolution in the phase space and the "stopping
 surface" where evolution of the material under the fixed conditions
 stops (see Sec.11 below).
\par
As $f \rightarrow 0$ and the kinetic processes are stopped, the
system tends to the "natural" limit state which determines the
relation between the "ground state energy" $F(S,\inel )$ and the
residual stresses (see Sec.8.3 below).\par

\subsection{Spacial subsystem.}
 The spatial part (7.7) of aging equations represents the
 system of PDE for the metric $g_{IJ}$ having the form

\begin{equation}
-\beta{\mathcal E}^{AB}(g)-S^{-1}S^{IA}Q^{BN}_{IM}(\xi_{p}^{2}g
)^{M}_{N}+W=S^{(AB)}.
\end{equation}
Here $Q^{BN}_{IM}=\frac{\partial^{2} \chi}{\partial
K^{I}_{B}\partial K^{M}_{N}}$ and $\xi_{p}=\partial_{t}-{\vec N}$ is
the principal part of the 1st order linear operator $\xi
=\partial_{t}-{\mathcal L}_{\vec N}$. The term $W$ in the left side
depends on the metric coefficients, function $S$ and their first
derivatives. Einstein tensor ${\mathcal E}(g)$ is linear by the
second-order space derivatives of $g$.  Thus, this system is
quasilinear evolutional second order system for metric $g$.  It can
be easily transformed to the normal form under simple conditions on
the dissipative potential $\chi$.\par

\subsection{Statical case.} Consider the case where ${\vec N }=0, U=0$,
$f$ does not depend on $G,S$ explicitly, $S,g$ are time-independent,
$\beta =const$. Then the system of aging equations is reduced to the
following form (here and below $\tF =SF$)\beq
\begin{cases}
\frac{\partial \tF}{\partial S}+\beta R(g)=f(\el ), \\
(S)^{-1}\frac{\partial \tF}{\partial g_{AB}}-\beta {\mathcal
E}^{AB}+(\rho_{0}S)^{-1}[\Delta_{g}(\rho_{0}S)g^{AB}+Hess_{g}^{AB}(\rho_{0}S)]+\frac{1}{2}(F+\beta
R(g)) g^{AB}=\frac{1}{2}(f g^{AB}-S^{(AB)}).
\end{cases}
\eeq In the absence of the strain energy, i.e. when $f(\el )=0,\
b^{(AB)}=0$ system (8.3) has the trivial solution $S=const,
g=g_{0}$.
\par
Calculate $S^{AB}$ through the Cauchy stress tensor using (6.3) as
follows: $S^{AB}=P^{A}_{j}\phi^{j}_{C}g^{CB}=(J(\phi)\phi^{-1\
A}_{s}h^{is}\sigma_{ij})\phi^{j}_{C}g^{CB}=J(\phi
)g^{BC}\phi^{j}_{C}\phi^{-1\ A}_{s}h^{is}\sigma_{ij}=
\frac{\sqrt{\vert h\vert}}{\sqrt{\vert g\vert}}g^{BC}\sigma^{A}_{C}.
$
\par
Multiplying the first equation in (8.3) by $\frac{1}{2}g^{AB}$ and
subtracting from the second we get

\beq [\frac{\partial F}{\partial g_{AB}}-\frac{1}{2}\frac{\partial
\tF}{\partial S}g^{AB}]\sqrt{\vert
g\vert}+(\rho_{0}S)^{-1}\sqrt{\vert
g\vert}[\Delta_{g}(\rho_{0}S)g^{AB}+Hess_{g}^{AB}(\rho_{0}S)]-\beta
{\mathcal E}^{AB}\sqrt{\vert g\vert}=-\frac{1}{2}\sqrt{\vert
h\vert}g^{(B\vert C}\sigma^{A)}_{C}. \eeq

This is the balance equation between the metric characteristics
(Einstein tensor, "ground state energy", lapse function $S$) and the
stresses in the body.  It is especially simple in the case where
$S\equiv 1$ is absent from $F$:

\beq \frac{\partial F}{\partial g_{AB}}\sqrt{\vert g\vert}-\beta
{\mathcal E}^{AB}\sqrt{\vert g\vert}+\rho_{0}^{-1}\sqrt{\vert
g\vert}[(\Delta_{g}\rho_{0})g^{AB}+Hess_{g}^{AB}\rho_{0}]=-\frac{1}{2}\sqrt{\vert
h\vert}g^{(B\vert C}\sigma^{A)}_{C}. \eeq
 Here we can see how the
curvature of material metric and the density of non-homogeneities
may be a source of the stresses in the body in the absence of
elastic deformation, i.e. when the conventional strain tensor
$E^{el\ con}=\frac{1}{2}ln(g_{0}^{-1}C_{3}(\phi))$ is zero.  Namely,
in such a case though the conventional strain tensor is zero,
decline of the Cauchy metric $C_{3}(\phi )$ from the material metric
$g$ is not zero.  Subsequently stress tensor $S$ is not zero.
Equation (8.5) thus describes the self-equilibrated stress resulting
from the curvature of the metric $g$ and is related to the
incompatibility of embedding of the solid into the physical space.
The first term on the left in (8.5) is related to the deviation of
the total energy from its stationary value.\par

One example of this situation a nonhomogeneous  chemical
transformation (oxidation) of material, which results in the
variation of material density and an incompatibility with the
reference configuration. A more specific example of stress induced
chemical transformation is discussed below in section 8.6.

\subsection{Almost flat case.} Here we use essentially that the
dimension of $B$ is 3. In the case, where $Ric(g_{t})\approx 0$,  a
good approximation of the general system (7.1-2) can be proposed. If
the total deformation $\phi $ is approximated by the "{\it ground
deformation}" $\bar{\phi }$ (i.e. deformation ${\bar{\phi }}(X,T)$
such that ${\bar{\phi }}^{*}h=g_{T}$, recall that this is the
synchronous case!) in the evaluation of the EMT $T^{IJ}$ on the
right side of aging equations (7.2), the latter becomes decoupled
from the equilibrium equations (7.1). This allows us to study the
aging equations separately from the elasticity equations and, after
obtaining solution for $G$, substitute them into the elastic
equilibrium equation (7.1) and solve it as the conventional
elasticity equation {\bf with variable elastic moduli}. \par

\subsection{ Homogeneous media}
\par In the case of a homogeneous material
(\cite{M}) metric $G$ depends on $T$ only, and Einstein tensor
${\tilde{\mathcal E}}_{IJ}(g)$ is identically zero. As a result,
(7.2) becomes a system of quasi-linear {\bf ordinary} differential
equations of the second order for the lapse function $N$ and the
material 3D metric $ g_{IJ}$. The Cauchy problem for this system is
correct under some mild conditions to the dissipative potential
$\chi$.\par
 The linearized version of aging  equations of 1D homogeneous rod was discussed in
(\cite{CP4}). In Sec.11 we shall briefly describe the study of some
aging problems for a homogeneous rod (more detailed presentation
will be published elsewhere, see \cite{CP3}).\par

We conclude this section with two model examples that show the type
of material behavior that can be studied using presented
approach.\par
\begin{example}{\bf  Modeling of Necking Phenomena in Polymers.} \par

Delayed Necking, observed in various engineering thermoplastics, is
a pictorial illustration of traveling wave solution. Necking in
general is a localized large deformation (drawing) of a polymer with
a distinct boundary between the drawn and undrawn material domains (
\cite{L,Z1, Z2}). Delayed necking takes place in a rod in uniaxial
tension, i.e., under constant applied load when the initial
Piola-Kirchoff stress $S_{11}(0)$  is less then the yield stress. At
first a uniform creep takes place, i.e., a uniform stretching with a
draw ratio $\lambda  = l/l_{0}$, where $l$ stands for an actual
(current) length scale.  After certain time interval when the
increasing stress $S_{11}$ reaches the yield stress value, necking,
also called "cold drawing" with a natural draw ratio $\lambda =
l_{1}/l_{0}$ starts, i.e., strain localization is formed and
propagates along the rod with a constant speed $N^1$. The observed
elongation results exclusively from a transformation of the original
material adjacent to the neck boundary into the drawn (oriented)
state and propagation of the boundary along the rod, as depicted in
Fig. 6.\par

 We consider here a 1D model of a rod, with the lapse function $S=1$ and
the shift vector field $\vec{N} =N^{1}\frac{\partial}{\partial X}$
being constant (see \cite{CP2} for a 3D model of the necking
process). Denote by $g=g_{11}(t,X)=\lambda^2 g_{0}$ the only
component of material metric. Take the "ground state" energy as
\[
F(\lambda )=(\lambda -\lambda_{0})^{2}(a+b(\lambda
-\lambda_{1})^{2})
\]
where the elongation of the rod $\lambda(t,X)
=\frac{1}{2}\frac{g}{g_{0}}$ is the drawing variable,
$\lambda_{0}=1,\lambda_{1}$ are two states (to compare with example
of the creep in Sec.11 put $\lambda =e^{\eta}\approx 1+\eta $). This
is the simplest function that admits two different stable states
(metrics) with equal chances when true stress reaches a critical
value.\par

The metric Lagrangian is reduced to $L_{m}=F(\lambda )+\chi(K),$
where $K=(g^{-1}u_{G}\cdot g)=2\partial_{\tau} \eta ,\ \eta=
ln(\lambda )$ where $\partial_{\tau}=u_{G}=\partial
_{t}-N^{1}\partial _{X}$.\par Experimental data suggest that in the
necking the material density variation is negligible, thus we take
$\rho_{0}(t,X)\lambda (t,X)=\rho_{0}(0,X)$. As a result, the action
takes the form
\begin{multline}
A(\lambda (t,X))=\int_{[0,t_{1}]\times
[0,L]}[F(\lambda)+\chi(2\partial_{\tau} ln(\lambda) )
 ]dt\wedge dX=\\ \int_{[0,t_{1}]\times [0,L]}[(\lambda -\lambda_{0})^{2}(a+b(\lambda
-\lambda_{1})^{2})- \frac{\mathcal
F}{A_{0}}(\lambda-\lambda_{0})^{2} +\chi(2\partial_{\tau}ln(\lambda)
)]dt\wedge dX,
\end{multline}
where $A_{0}$ is the initial cross section of the rod and $\mathcal
F$ is the force acting on the right end pulling in $X$-direction.
The second term here represents the work of the load on non-elastic
deformation. Consider the case where $\mathcal F$ is large enough to
change sign of the quadratic part of the"ground energy" $F$. For
simplicity we take $\frac{\mathcal F}{A_{0}}=a $. \par

Variation by $\lambda $ leads to the aging equation in the form

\[
\chi'' \lambda_{\tau
\tau}-\frac{\lambda_{\tau}}{2}[\chi'+(2\lambda^{-1}\lambda_{\tau})\chi'']-\frac{\lambda^2}{4}{\bar
F}'(\lambda )=0,
\]
where ${\bar F}(\lambda )=b(\lambda -\lambda_{0})^{2}(\lambda
-\lambda_{1})^{2}.$

The 2D dynamical system corresponding to this equation has
equilibria points $(\lambda_{i},0),\ i=0,1,2$ at the roots
$\lambda_{0},\lambda_{2}=\frac{1}{2}(\lambda_{0}+\lambda_{1}),\lambda_{1})$
of the polynomial $F'(\lambda).$  If the dissipative potential $\chi
(u)$ satisfies to the conditions $\chi'(0)=0,\chi''(0)>0$, root
$\lambda_{2}$ is the center while other two are saddles whose
separatrix loop enclose the elliptic region.\par
 For a
given $F(g)$, when the stress reaches the initiation level and from
the trivial solution for $g$ there bifurcates the separatrix
solution, then, we get the "traveling wave solutions" in the form of
"kink" (\cite{CP2}), propagating with the the speed $N^1$ along the
rod, for the metric $g(X,t)$ (and, by the mass conservation law, for
the density $\rho _{0}(X,t)$).
\begin{figure}
\scalebox{.70}{\includegraphics{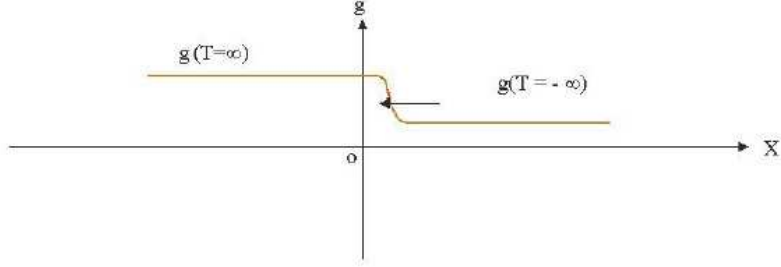}} \caption{Necking of
a 1D Rod}
\end{figure}
\end{example}

\subsection{Example: Variation of Material Metric $g$ due to the Chemical Degradation}
\par
Here the evolution of the uniform material metric to the piecewise
constant metric with the jump along the interface between a layer of
chemically degraded material  and the original material follows the
kinetics of chemical degradation (see \cite{CZCSSB}).\par

Consider a thin-walled thermoplastic tubing employed for transport
of chemically aggressive fluid.  In time, the inner surface layer of
material undergoes chemical degradation due to interaction with
aggressive fluid flow.\par

Chemical degradation is manifested in an increase of the material
density $\rho_{0}$, significant reduction in toughness (resistance
to cracking) and a subtle change in yield strength, Young's modulus
and other thermo-mechanical properties.\par

Assuming the homogeneity of degraded layer we see that the original
euclidian material reference metric in degraded ring evolves (see
the mass conservation law (3.4)) which generates a jump on the
interface with the outer layer of unchanged material.  Continuity of
normal stresses on the interface allows us to describe the final
state of the system by elementary methods presented below.\par

Consider a thin ring (see Figure 7) which represent the 2D
cross-section of the tubing.  The wall thickness $t=R_{o}-R_{i}$ is
small in comparison to the outer radius $R_{o}$: $t/R_{o}\ll  1$.
$R_{d}$ in Fig ** stands for the radius of interface between the
layer of degraded material and unchanged layer. The depth of
degradation $t_{d}=R_{d}-R_{i}$ is relatively small: $t_{d}/t \ll
1$.\par

Select the polar coordinate system $(r,\theta )$.  2D material
metrics of the initial ($g^{0}$) and degraded ($g'$) states are
\[
g^{0}=\left( \begin{smallmatrix} 1 & 0\\ 0 & r(0)^{2}
\end{smallmatrix}\right) ,\ \vert g^{0}\vert =r^{2}(0);\ g'=\left( \begin{smallmatrix}
1+\epsilon  & 0\\ 0 & r^{'2}
\end{smallmatrix}\right),\ \vert g'\vert =(1+\epsilon)r^{'2},
\]
where $r'=(1+\epsilon)r$ and $\epsilon$ is a small variation of
scale in the radial direction.\par

Mass conservation law $\rho_{0}\sqrt{\vert g^{0}\vert }=\rho_{0}'
\sqrt{\vert g'\vert }$ relates density variation
$\rho'_{0}=\rho_{0}+\Delta \rho_{0},\ \frac{\Delta
\rho_{0}}{\rho_{0}}\sim 10^{-3}$ with the change in material metric
\beq \frac{\rho_{0}}{\rho_{0}+\Delta
\rho_{0}}=\sqrt{\frac{(1+\epsilon)r^{'2}}{r(0)^2}}=(1+\epsilon)^{3/2}
\Longrightarrow 1-\frac{\Delta \rho_{0}}{\rho_{0}}\approx
1+\frac{3}{2}\epsilon +O(\epsilon^2). \eeq Therefore
$
\epsilon \approx -\frac{2}{3}\frac{\Delta
\rho_{0}}{\rho_{0}}+O(\epsilon^2).
$

Thus, the densification (i.e. $\Delta \rho_{0}>0$) leads to the
shrinkage of the thin ring of degraded material. If we remove the
constrains on shrinkage applied by the outer ring of original
material the gap \beq
 w=R_{d}^{0}-(1-\frac{2}{3}\frac{\Delta
\rho_{0}}{\rho_{0}})R^{0}_{d}= \frac{2}{3}\frac{\Delta
\rho_{0}}{\rho_{0}}R^{0}_{d}
\eeq
appears.\par

As a result of such constrains, the degraded material should be
elastically stretched to close the gap $w$.  This elastic
deformation has the form
\[
\phi (r, \theta)=\begin{cases}
r''(r)=(1+\frac{2}{3}\frac{\Delta\rho_{0}}{\rho_{0}})r',\\
\theta''=\theta'=\theta .
\end{cases}
\]
Under such a deformation elastic strain tensor
$E^{el}=\frac{1}{2}g^{-1}(C_{3}(\phi )-g), C_{3}(\phi
)=h_{ij}\phi^{i}_{,I}\phi^{j}_{,J}$ has the form
\beq E^{el}\approx
\frac{2}{3}\frac{\Delta\rho_{0}}{\rho_{0}}\begin{pmatrix} 1 & 0\\ 0
& 1 \end{pmatrix}. \eeq

The tensile strain (8.9) is directly translated into the tensile
radial stress via Hooke's law
\[
\sigma_{rr}=\frac{Y}{1-\nu}\frac{2}{3}\frac{\Delta\rho_{0}}{\rho_{0}}.
\]
Although hoop stresses $\sigma_{r\theta}$ may be discontinuous, the
equilibrium conditions requires continuity of radial stress across
the interface, ,
\[
\sigma_{rr}n_{r}\vert_{r=R_{d}-0}=\sigma_{rr}n_{r}\vert_{r=R_{d}+0}.
 \]
This implies that the outer ring of original material experiences
compressive stresses while the inner degraded layer is under
tension. The elastic strains $E^{el}$ jointly close the gap $w$ and
restore the compatibility of the Cauchy metric $g_{final}$ in the
whole domain:
\[
g_{final}=g'+E^{el}.
\]
Therefore while the material metric $g'$ has the jump leading to the
nonzero singular curvature along the interface, the final metric is
continuous and flat.

\begin{figure}
\scalebox{.80}{\includegraphics{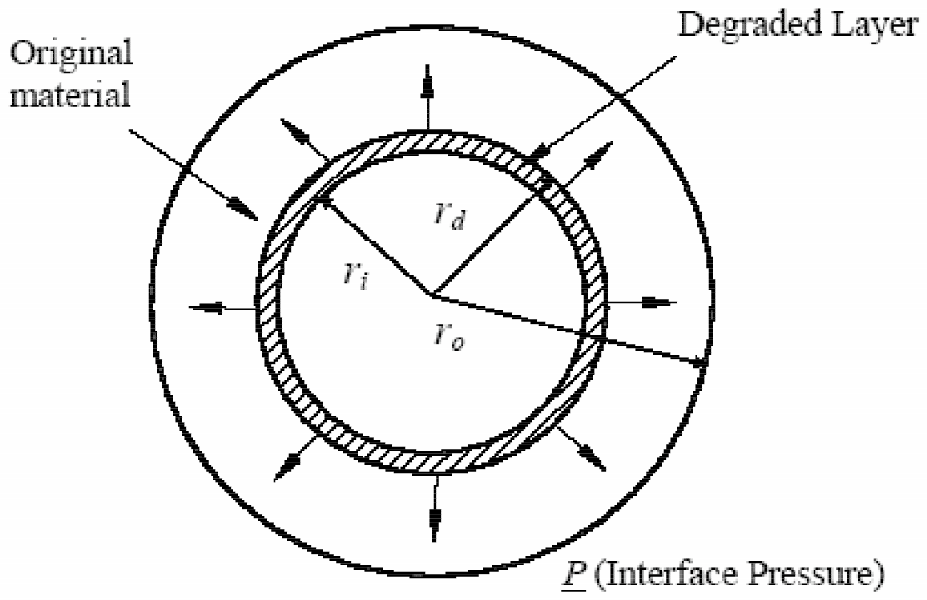}} \caption{A sketch of
polymer tubing cross section with inner degraded layer}
\end{figure}

\section{Physical and Material Balance Laws}
As it is typical for a Lagrangian Field Theory, action of any
one-parameter group of transformations of the space $P\times M$,
commuting with the projection to $P$, leads to the corresponding
balance law (See \cite{MH}). In particular, translations in the
"physical" space-time $M$ lead to the dynamical equations (7.1-2),
rotations in $M$ lead to the angular momentum balance law
(conservation law in the absence of applied torque). Respectively,
translations in the "material space-time" $P$ lead to the {\bf
energy balance law} (translations along the time $T$ axis) and to
the {\bf material momentum balance law} ("pseudomomentum" balance,
(\cite{M}, \cite{G}, \cite{H}), rotations in the material space $B$
lead to the "material angular momentum" balance law (\cite{M}).
\par
In the table below we present basic balance laws together with the
transformations generating them.  It is instructive to compare the
space and material balance laws as it has been considered
previously by several authors (\cite{G}, \cite{H}).

\begin{table}[h]
\begin{center}
\caption{Space and Material Balance Laws}
\begin{tabular}{| c | c | c | }\hline
Symmetry & \parbox[c]{4.7cm}{Physical space-time\\ (Material
independent)} &
 \parbox[c]{4.7cm}{Material space-time\\ (Space independent) } \\ \hline
Homogeneity of 3D-space  &
\parbox[c]{4.7cm}{ {\small Linear momentum balance law}\\
{\small (Equilibrium equations)}\\
$div(\sigma )= f$ } &
\parbox[c]{4.7cm}{ {\small Material momentum (pseudo-}\\ {\small (-momentum) balance law } \\
$\textbf{div(b)=fmat}$ }\\ \hline Time homogeneity &
\parbox[c]{4.7cm}{\small Energy balance law:\\ $\partial _{t}{\mathcal E}^{tot}=div({\mathcal
P}^{tot} $ )} & {\small Energy balance law} \\ \hline
Isotropy of 3D-space & \parbox[c]{4.7cm}{ {\small Angular momentum balance law } \\
{\small  $\equiv $ h-symmetry of Cauchy stress } \\ {\small
tensor} $\sigma$:      $\ \ \ \mathbf{I:\sigma =\sigma :I}$} &
\parbox[c]{4.7cm}{{\small Material angular momentum}\\
{\small balance law} $\equiv $ $C$-{\small symmetry of}\\
{\small Eshelby stress tensor} {\textbf b} \\
 $\mathbf{b:C=C:b} $ } \\ \hline
\end{tabular}
\end{center}
\end{table}

 Space and Material balance
(conservation) laws are related via the deformation gradient $d\phi
$.  Restricting ourselves to the synchronized case and writing the
 material balance laws in the form $\eta _{I}=0$ and their "physical"
 counterparts in the form $\nu_{i}=0$ we get the relationsship between these families
 of balance laws
\begin{equation}
\left(
\begin{array}{c}  \eta _{0} \\ \ldots \\ \ \eta _{4} \end{array}
\right)=\left(
\begin{array}{cccc} 1&\phi ^{1}_{,0} & \ldots &\phi ^{3}_{,0}\\
1&\phi ^{1}_{,1} & \ldots &\phi ^{3}_{,1}\\
\ldots & \ldots & \ldots &\ldots \\
0& \ldots &\ldots &\phi ^{3}_{,3}
\end{array}
\right) \cdot \left( \begin{array}{c}  \nu _{0}\\ \ldots \\ \nu
_{4}
\end{array} \right)
\end{equation}
  Similar to the case of
relativistic elasticity (\cite{KM}), the system of material balance
laws $\eta _{I}=0,\ I=1,2,3$ is equivalent to the elasticity
equations $\nu _{i}=0,\ i=1,2,3$, while the energy balance law $\eta
_{0}=0$ (which here is the {\bf material} conservation law as well
as the {\bf physical one}: in the case of synchronized history of
deformation $\phi $ material time $T$ and physical time $t$
coincide). As a result, the energy conservation law is the
consequence of the time translation invariance in both senses and
follows from any of these two systems: $\eta _{0}=\sum
_{i=1}^{i=3}\phi ^{i}_{,0}\nu _{i}.$ This reflects the fact that the
deformation we consider here are not truly 4-dimensional.\par
Balance laws (with the source terms) can be transformed into
conservation laws by adding new dynamical variables. In the theory
of uniform materials (\cite{M,EM,TN}) it is zero curvature
connection in the frame bundle over $M$ that is added to the list of
conventional dynamical variables, in our scheme - it is the 4D
material metric $G$.
\par

\section{Energy-Momentum Balance Law and the Eshelby Tensor.}

In this section we consider the Energy-Momentum balance law
resulting from the Least Action Principle and the space-time
symmetries. \par

Consider local rigid translations in the material space-time $X^{J}
\longmapsto X^{J}+\delta X^{J}.$ They generate a variation of
components $\phi ^{i}$ of the deformations, components $G_{IJ}$ of
material metric and their derivatives (we follow the arguments of
J.Eshelby (\cite{E3}).\par
  Taking the variation of the
Lagrangian density ${\mathcal L}=\sqrt{\vert G\vert }L=\sqrt{\vert
G\vert }(L_{g}(G)+L_{e}(G,E^{el}))$ with respect to the material
coordinates $X^J$, one obtains
\begin{eqnarray}
\frac{\delta {\mathcal L}}{\delta X^{J}} &=&
\sum_{i=0}^{i=3}\frac{\delta {\mathcal L}}{\delta \phi ^{i}}\phi
^{i}_{,J}+\frac{\partial }{\partial X^{I}} (\sum
_{i=1}^{i=3}\frac{\partial {\mathcal L}}{\partial \phi
^{i}_{,I}}\phi
^{i}_{,J})+ \frac{\delta {\mathcal L}}{\delta G^{AB} }G^{AB}_{,J}+  \nonumber  \\
& &\frac{\partial }{\partial X^{I}}\left( {\mathcal
E}(G)^{I}_{J}=\frac{\partial {\mathcal L}}{\partial
G^{AB}_{,I}}G^{AB}_{,J}+ \frac{\partial {\mathcal L}}{\partial
G^{AB}_{,IK}}G^{AB}_{,KJ}-\frac{\partial }{\partial X^{K}}\left(
\frac{\partial {\mathcal L}}{\partial G^{AB}_{,IK}}\right)
G^{AB}_{,J}\right).\hskip0.8cm
\end{eqnarray}
 The last term in the right side of (10.1) includes a definition of
the (1,1)-tensor density ${\mathcal E}(G)$. Employing the
Euler-Lagrange equations (7.1-2), we obtain for the Total
Energy-Momentum Tensor (density) \begin{equation} {\mathcal
E}^{tot}= -{\mathcal L}\delta ^{I}_{J}-S^{I}_{J}+ {\mathcal
E}(G)^{I}_{J},
\end{equation}
the conservation law
\begin{equation}
div_{G_{0}}({\mathcal E}^{tot})=\frac{\partial }{\partial
X^{I}}{\mathcal E}^{tot\ I}_{J}=0,\ J=0,1,2,3.
\end{equation}
Divergence here is taken with respect to the 4D "reference" metric
$G_{0}$. Since $L_{m}$ does not depend on deformation $\phi$ and the
body forces potential $U$ does not depend on its derivatives while
$L_{e}=-\rho_{0}f-\rho_{0}U$,
\begin{equation}
{\mathcal S}^{I}_{J}=-\sum _{i=1}^{i=3}\frac{\partial {\mathcal
L}}{\partial \phi ^{i}_{,I}}\phi ^{i}_{,J}=\sum
_{i=1}^{i=3}P^{I}_{i}\phi ^{i}_{J}S\sqrt{\vert g\vert },
\end{equation}
which is the 4D-version of the (density of) Second Piola-Kirchoff
Stress Tensor.
\par

Rewrite ${\mathcal E}^{tot}$ in the form: ${\mathcal
E}^{tot}={\mathcal B}+U\delta^{I}_{J}\sqrt{\Vert G\Vert} +{\mathcal
L}_{m}\delta^{I}_{J}+{\mathcal E}(G)$ where we denoted by $\mathcal
B$ the tensor density ${\mathcal B}=b\sqrt{\vert G\vert}$ of the
Eshelby EM Tensor. Then the equation (10.3) takes the form \beq
div_{G_{0}}{\mathcal B}={\mathcal
B}^{I}_{J,I}=-div_{G_{0}}(U\delta^{I}_{J}\sqrt{\Vert G\Vert}
+{\mathcal L}_{m}\delta^{I}_{J}+{\mathcal E}(G)), \eeq where in the
right side only metrical quantities and the potential $U$ of the
body forces are left.\par

The second term and the metrical part of the third term in the right
side of (10.5) are related to the ground state of the Lagrangian
density i.e. to the inhomogeneity of "cohesive energy" and the
"material flows". The elastic part of the third term on the right is
related to a variation of elastic moduli if these moduli depend on
the derivatives of the metric $G$. Equality (10.5) can be easily
rewritten in terms of covariant derivatives with respect to the
metric $G$.\par

 Taking $J=0$ in (10.6) we arrive at the energy conservation law in ADM notations
 (using $t$ instead of $X^{0}$)
\begin{equation}
\frac{\partial }{\partial t}((f+U+L_{m})S\sqrt{ \g}+{\mathcal
E}^{0}_{0}) = \sum _{I=1}^{I=3}\frac{\partial }{\partial X^{I}}
\left( \sum _{i=1}^{i=3}P^{I}_{i}\phi ^{i}_{,0}S\sqrt{ \g}-
{\mathcal E}(G)^{I}_{0}. \right)
\end{equation}
\par

  Equation (10.6) has the form $
\frac{\partial ({\mbox Total Energy Density})}{\partial T}={\mbox
Total  Flow Density}, $ with the total (inner) energy density given
by
\begin{equation}
{\mathcal E}^{tot \ 0}_{0}=(f+U+L_{m})S\sqrt{ \g}+{\mathcal
E}(G)^{0}_{0}.
\end{equation}
   The total energy is the sum of the following parts: elastic
energy $f$, potential energy of the volume forces $U$ , cohesive
"ground state" energy  - the term $F(E^{in} ,S)$ in $L_{m}$,
inhomogeneities energy from the curvature density $\beta
R(g)S\sqrt{\g }$ and corresponding terms of ${\mathcal E}^{0}_{0}$,
"kinetic metric energy" that is defined by the term s produced by
$\chi (K)$ in $L_{m}$  and ${\mathcal E}^{0}_{0}$ and reflects the
intensity of irreversible deformation and "metrical volume change
energy" coming from the $div_{g}({\vec N})$-terms.\par
  The
sum on the right side of (10.5) consists of the flow of the
Piola-Kirchoff stress tensor density $\sum _{I=1}^{I=3}
[P^{I}_{i}\phi ^{i}_{,0}S\sqrt{ \g }]_{,I}$ and the flows related to
the change of the material metric - internal material flows, flows
of inhomogeneities (coming from the curvature $R(g)$ etc.
 \par

If the metric $G$ does not depend on time (i.e. ${\bar N}=0,K=0,\
G=G_{0}$) and if $Ric(g_{t})=0$, one obtains the conventional energy
conservation law of Elasticity Theory (\cite{MH}, Chapter 5, Sec.5):
$\frac{\partial (f+U)}{\partial T}=-\sum _{I=1}^{I=3}\frac{\partial
}{\partial X^{I}} (P^{I}_{i}\phi ^{i}_{,0}),$.\par

 \begin{example}[Block diagonal metric $G$, synchronous deformation and homogeneous
media]
\par
In this case we have $Ric(g_{t})=0,\ {\tilde N}=0,g=g(t),S=S(t)$,
the extrinsic
curvature has the form $ K^{I}_{J}=\left(\begin{array}{cc} 0&0\\
0& S^{-1}g^{IK}g_{KJ,0} \end{array}\right) $ and $\chi (K)$ is the
only term in the Lagrangian containing time derivatives.
\par
 In addition to this, no flow terms except the
usual Piola-Kirchoff flow appear on the right side in the energy
balance law which takes the form
\begin{equation}
\frac{\partial }{\partial t}\left( (f+U+L_{m})S\sqrt{\g
}+S^{-1}g_{AB,0}\left[\frac{\partial \chi
}{K^{M}_{B}}g^{MA}+\frac{\partial \chi }{ K^{M}_{A}} g^{MB}\right]
S\sqrt{\g } \right) =+\sum _{I=1}^{I=3}(P^{I}_{i}\phi
^{i}_{,0}S\sqrt{\vert g\vert })_{,I}.
\end{equation}
This equation describes how the energy supplied by the boundary load
spreads not just to the increase of the strain energy, but also to
the change of its "cohesive energy" of the material ($F(S, \vert
g\vert)$) and to the acceleration of the aging processes.
\end{example}

\section{Aging of a homogeneous rod}
In considering three types of inelastic processes in a tensile
homogeneous rod: unconstrained aging, stress relaxation and creep
(see \cite{CP3} for more detailed exposition) we assume that ${\vec
N}=0, R(g)=0$, $S(t=0)=1$ and that $S(t)$ is increasing to a certain
level depending on the initial state of the body and the
process.\par

\subsection{Deformation, strain tensors and tensor $K$}
Introduce material cylindrical coordinates $(R,\Theta ,Z)$ in the
reference state of a rod $B$. Spacial cylindrical coordinates
$(r,\theta ,z)$ are introduced in the physical space $R^3$.  In
addition we normalize the initial state $g(0)$ of material metric
taking $g(0)=g_{0}$. \par

We consider the class of time dependent (total) deformations $\phi $
of the from \beq \phi_{t}:(R,\Theta ,Z)\rightarrow (r=
\mu(t,Z)R,\theta =\Theta ,Z=k(Z,t)), \eeq
 with  $\mu, k$ representing amount of "stretch" in radial and axial
 directions respectively. \par Material metric
$g_{t}$ is flat (homogeneous case!) and is generated by a global
deformation $\phi_{m}$ of the same type as (11.1), with
$\mu_{m},k_{m}$ the same as above: $g_{t}=\phi_{m}(t,\cdot )^{*}h$.
As a result

\beq
 g=\phi_{m}^{*}h=\left( \begin{smallmatrix} \mu_{m}^{2} & 0 & R \mu_{m}
\mu_{m,Z}\\
0 &  R^{2}\mu^{2}_{m} & 0\\
R \mu_{m}\mu_{m,Z} & 0 & \lambda ^{2}+R^{2} \mu^{2}_{m,Z}
\end{smallmatrix}\right) ,\ C(\phi)=\left( \begin{smallmatrix} { \mu}^{ 2} & 0 & R{
\mu}{\mu}_{,Z}\\
0 & R^{2}{\mu}^{2} & 0\\
R{ \mu}{ \mu}_{,Z} & 0 & \lambda ^{2}+R^{2}{ \mu}^{ 2}_{,Z}
\end{smallmatrix} \right),
 \eeq
where $\lambda =k_{,Z},\lambda_{m}=k_{m,Z}$. For a homogeneous rod
 $k(Z,t)=\lambda(t)Z,\ k_{m}(Z,t)=\lambda_{m}(t)Z$.\par

In the elasticity theory (see \cite{GZ}) it is customary to present
deformation (total and inelastic as well) as the composition of a
uniform dilatation with the axial expansion factor $\lambda_{v}(t)$
and of the volume preserving normal expansion with the factor
$\lambda_{d}$: $\lambda =\lambda_{v}\lambda_{d}$. We will obtain
$\mu =\lambda_{v}\lambda_{d}^{-1/2}$ and $\sqrt{\g
}=\lambda_{v}^{3}R$ (and the same for
$\lambda_{mv},\lambda_{md}$).\par

As a result, the inelastic strain tensor can be written in the
following form

\beq E^{in}=\frac{1}{2}(g_{0}^{-1}g)=\left( \begin{smallmatrix}  ln(\mu) & 0 & 0\\
0 & ln(\mu ) & 0 \\
0 & 0 & ln(\lambda )\end{smallmatrix}\right) = \left( \begin{smallmatrix} \xi -\frac{1}{2}\eta & 0 & 0\\
0 & \xi -\frac{1}{2}\eta & 0\\
0 & 0 &\xi +\eta  \end{smallmatrix}\right) \eeq

Here we introduced the variables $\xi =ln(\lambda_{m\ v}), \eta
=ln(\lambda_{m\ d})$.  As a result, calculating basic invariants of
these tensors we see that the "ground state" energy $F$ is the
function of $S,\xi ,\eta^{2}.$\par

 Decomposing the total deformation as the
composition of inelastic and elastic one and assuming that elastic
deformation is small compared to $1$ we write:
\[
\lambda_{v}=\lambda_{v\ m}(1+\epsilon_{v}),\ \lambda_{d}=\lambda_{d\
m}(1+\epsilon_{d}).
\]
In these notations elastic strain tensor takes the conventional
diagonal form
\[
E^{el}=\frac{1}{2}ln(g^{-1}C({\phi^{tot}}))\approx diag(\
\epsilon_{v}-\frac{1}{2}\epsilon_{d},
\epsilon_{v}-\frac{1}{2}\epsilon_{d}, \epsilon_{v}+\epsilon_{d} ).
\]
 Strain energy for our (homogeneous) rod will now take the form

\beq f=\frac{K}{2}\epsilon_{v}^{2}+\frac{3\mu }{2}\epsilon_{d}^{2}
\eeq
with the bulk coefficient $K$ and the Lame coefficient
$\mu$.\par

 Mass conservation law (3.4) takes the form
$\rho_{0}(t)=\lambda_{v}^{3}\rho_{0}(0).$\par

The spacial part of the tensor $K$ for a homogeneous rod takes the
diagonal form

\beq
 K=S^{-1}diag(
2\xi_{t}-\eta_{t},
 2\xi_{t}+2\eta_{t},
2\xi_{t}-\eta_{t} ).
 \eeq
Thus, $Tr(K)=6S^{-1}\xi_{t},\
Tr(K-\frac{1}{3}Tr(K)I)^2=6S^{-2}\eta_{t}^2 $ and dissipative
potential $\chi (K)$ is the function of arguments $S^{-1}\xi_{t},
S^{-1}\eta_{t}$.\par
\begin{remark} In general, aging equation for the described
situation have the form of a 3D degenerate Lagrangian system (we
refer to \cite{CP3}, or \cite{P} for more details).  In the cases of
the processes studied below this system reduces to the 2D degenerate
dynamical system.  In all three cases one can trivially solve
elasticity equations, exclude elastic variables
$\epsilon_{v},\epsilon_{d}$ from aging equations and, therefore, to
close the system of aging equations.
\end{remark}

\subsection{Unconstrained aging.}
Unconstrained aging (shortly UA) is the simplest example of a
material evolution. A sample of material (rod) is prepared and then
is left without any constraints or load applied to it. Usually the process
 of aging is manifested in a variation of material density, or a specific volume
 change  up to a saturation point, when the observable evolution stops.
 In many polymers the aging is accompanied by shrinkage up to a few percent
  of initial volume.  This
diminishing in volume (2-5\%) is called {\bf unconstrained aging}.
We discuss here a model for UA in terms of variables $(S,\xi)$
(dilatational deformation plays negligible role in UA). No strain
energy is present, stress is zero.\par
  We take the "ground state energy" to be
\[ F_{UA}(S,\xi )=
(c_{1}+c_{2}S+(p\xi S^{-1}+k\xi ^2) \] with
$k>0,p<0,c_{1}<0,c_{2}>0$ and the dissipative potential
\[
\chi_{UA}(K )=\alpha (S^{-1}\xi_{t})^2.
\]
Integrating over the volume of the rod we get the action in the form

\begin{equation}
A(\xi ,S)=V\int_{0}^{T}((c_{1}S+c_{2}S^{2}+(p\xi +k\xi^{2}S)+\alpha
S(S^{-1}\xi_{t})^{2})dt.
\end{equation}
Euler-Lagrange Equations of UA can be reduced to the following
dynamical system
\begin{equation}
\begin{cases}
\xi_{t}&=-S\left( \frac{c_{1}+2c_{2}S+k\xi^2 }{\alpha } \right)^{\frac{1}{2}},\\
S_{t}&=-\frac{p}{2c_{2}}\left( \frac{c_{1}+2c_{2}S+k\xi^2 }{\alpha }
\right)^{\frac{1}{2}}.
\end{cases}
\end{equation}
We have here $\xi_{t}\leqq 0, S_{t}\geqq 0$.\par

 Equation (7.5) takes here the form $ \alpha
\xi_{t}^{2}=S^{2}\frac{\partial (\tF)}{\partial S} ,$  where $\tF
=SF$.\par
 Take $\alpha
=-1$. Then, the domain of admissible dynamics
 defined by the positivity of expression under the
square root is
\begin{equation}
S\leqq -\frac{1}{2c_{2}}(c_{1}+k\xi^{2} ),
\end{equation}
and the curve where evolution stops when the phase trajectory
reaches the final state is $ S= -\frac{1}{2c_{2}}(c_{1}+k\xi^2 ).
$\par
 The "ground state energy" $F$ is
negative at initial moment and that it increases during the
evolution.\par

 System (11.7) has the first integral
$J=c_{2}S^2 -p\xi .$ Choosing an initial point $(S(0),\xi(0)=0)$ of
a trajectory $\Gamma$ in the domain of admissible dynamics. Along
$\Gamma $ we have $J(\xi ,S)=J(S(0),0)$ If we calculate $S(t)$ as
the function of $\xi (t)$ along $\Gamma $, substitute into the first
equation (11.7) and separate variables in this equation we get the
$\xi(t)$ as the explicit function of parameters of the problem and
initial value $S(0)$ in terms of elliptic functions (see
\cite{CP3}).\par

Figure 8 shows a family of shrinkage curves corresponding to the
various values of $S(0)=1,1.3,1.6,1.9$ (which represent the initial
aging of the material). Apparently, the higher is the initial age,
the less shrinkage is observed.
\begin{figure}
\scalebox{.50}{\includegraphics{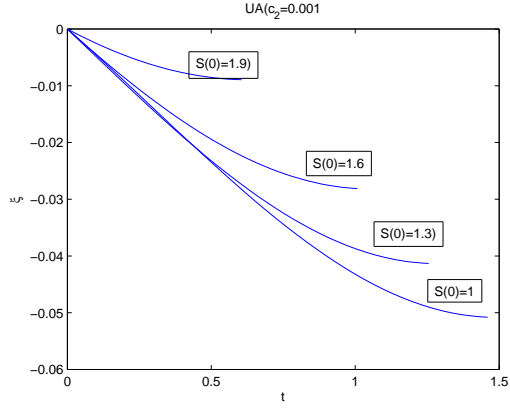}} \caption{Aging curves}
\end{figure}

\vskip0.5cm

When a load applied to the rod reaches certain level, new processes
may start. These new processes (going on the background of the UA)
initiate action of a new part of the "ground state" $F(S,\xi ,\eta
)$ and activates the new kinetic potential $\chi_{2}(K)$. For the
description of stress relaxation and creep we choose the dissipative
potential $\chi_{2}(K)=\frac{K}{\beta
D}ln(\frac{K}{D})-\frac{1}{\beta}(1+\frac{K}{D})ln(1+\frac{K}{D})$
corresponding to the phenomenological Dorn relation between the
stress and the strain rate $\dot \eta $ (see \cite{BS}, Sec.2.3 and
the footnote in Sec.5 above). Unconstrained aging is much slower and
leads to smaller changes then both stress relaxation and the creep.
That is why we may with good accuracy disregard the UA while
describing two other processes.

\subsection{Stress relaxation}

In the case of a {\bf stress relaxation} (SR) we fix the rod of
initial length $L$ at the left end and then quickly pull (or
compress) it uniaxially and fast (elastically) until it reach
certain length $L^{*}.$ Then we fix right end as well, leaving side
surface of the rod free. In this configuration the only component of
Cauchy stress that is nonzero is $\sigma_{zz}$. For the SR the
volume change is negligible and we have $ \lambda_{m\ v}=1,
\lambda_{m}=\lambda_{m\ d}$\par

Initially all the stretching is due to elastic process and $
\lambda^{*}_{d}=L^{*}/L=(1+\epsilon_{z}(0)).$ Then the inelastic
deformation starts to increase in expense of elastic one maintaining
the total strain constant. The reduction of elastic strain is
directly translated into the reduction of stresses via Hooke's law.
 The total elongation at moment $t$ can be decomposed as follows
\begin{equation}
\lambda^{*}_{d}=(1+\epsilon_{z}(t))\lambda_{m,d}(t)=(1+\epsilon_{z}(t))e^{\eta
(t)}
\end{equation}
and therefore $ \epsilon_{z}(t)= \eta^{*}-\eta (t), $ where
$\eta^{*}=ln(\lambda^{*}). $

From Hooke's law $ \sigma_{zz}=Y\epsilon_{z},$ where $Y$ is the
Young module. Thus for the strain energy expression we obtain
\begin{equation}
f(\eta
)=\frac{1}{2}\sigma_{zz}\epsilon_{z}=\frac{1}{2}Y\epsilon_{z}^{2}=\frac{Y}{2}(\eta^{*}-\eta
(t))^{2}.
\end{equation}
For pure stress relaxation (without background UA)
\[F=F_{SR}(S,\eta)=(q_{1}+q_{2}S) + \eta
(b_{0}S^{-1}+b_{1}+a_{1}\eta ),\ q_{1}<0,q_{2}<0,b_{0}<0,a_{1}<0,
\]
with the coefficients different from those of the slow UA. \par

Action $A(S(t), \eta(t) )$ now takes the form
\begin{multline}
A(S,\eta )=\int_{0}^{T}\left[ (q_{1}+q_{2}S) + \eta
(b_{0}S^{-1}+b_{1}+a_{1}\eta )+\frac{Y}{2}(\eta^{*}-\eta
(t))^{2}+S\chi (S^{-1}\eta_{t})\right]Sdt=\\
=\int_{0}^{T}\left[ {\tilde F}_{SR}(S,\eta)+ S\chi(S^{-1}\eta_{t})
\right] dt,
\end{multline}
where
\begin{multline}
{\tilde F}_{SR}(S,\eta)=SF_{SR}=(q_{1}S+q_{2}S^2 ) + \eta
(b_{0}+b_{1}S+a_{1}\eta S)+S\frac{Y}{2}(\eta^{*}-\eta (t))^{2}=\\
 q_{2}S^2 +p_{2}(\eta )S,
\end{multline}
where $ p_{2}(\eta)=(q_{1}+\frac{Y}{2}\eta^{*\ 2})+(b_{1}-Y
\eta^{*})\eta +(a_{1}+\frac{Y}{2})\eta^2 . $

We have ${\tilde F}_{S}=2q_{2}S+p_{2}(\eta )$ and the domain of
admissible motion is defined by
\[
D_{ad}=\{ (\eta ,S)\vert \eta >0, S\geqq 1,\ S<-\frac{p_{2}(\eta
)}{2q_{2}}\},
\]
while the stoping curve has the form $ S=-\frac{p_{2}(\eta
)}{2q_{2}}. $\par Aging equations (7.5-7.7) will now take the form

\begin{equation}
\begin{cases}
\eta_{t}&=S\psi^{-1}(2q_{2}S+p_{2}(\eta))= S\psi^{-1}(
2q_{2}S+[(q_{1}+\frac{Y}{2}\eta^{*\ 2})+(b_{1}-Y \eta^{*})\eta
+(a_{1}+\frac{Y}{2})\eta^2 ]),\\
S_{t}&=\frac{b_{0}}{2q_{2}}\psi^{-1}(2q_{2}S+p_{2}(\eta))=
S\psi^{-1}( 2q_{2}S+[(q_{1}+\frac{Y}{2}\eta^{*\ 2})+(b_{1}-Y
\eta^{*})\eta +(a_{1}+\frac{Y}{2})\eta^2 ]).
\end{cases}
\end{equation}

This system has the first integral $ J=b_{0}\eta -q_{2}S^2,$ and the
phase trajectory corresponding to the initial value
$S=S(0),\eta(0)=0$ has the form
$S=\sqrt{S(0)^2+\frac{b_{0}}{q_{2}}\eta }. $ Using this one can find
analytic solutions $\eta(t)$ in terms of elliptic functions
(\cite{CP3}).\par
 On the Figure 9 we present results of calculations of the
stress relaxation $\sigma_{zz}(t)$ for several values of initial
stretching $L=0.1,0.15,0.2,0.5$ and for realistic values of
parameters of the problem. Values of $\sigma_{zz}(t)$ are found by
solving numerically system (11.13) for $\eta(t)$, calculating
elastic strain $\epsilon_{z}(t)$ and using the Hooke's law.
Apparently, the higher is the value of initial stretching, the
sharper is the stress relaxation curve and the higher is the
asymptotic value of stress.

\begin{figure}
\scalebox{.50}{\includegraphics{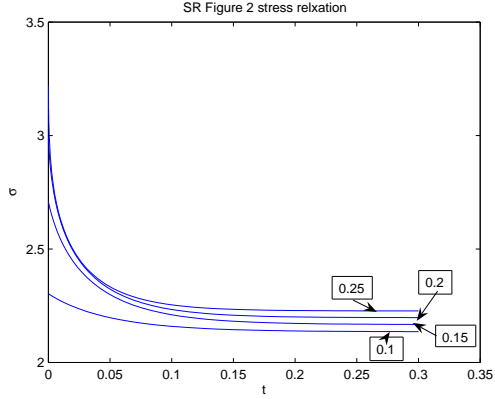}} \caption{Stress Relaxation:
Stress $\sigma_{zz}(t).$ for $\eta^{*}=0.1,0.15,0.2,0.25$.}
\end{figure}

\subsection{Creep}
In the case of the {\bf creep} we fix the left end of the rod and
apply force $\mathcal F$ in $Z$-direction to its right end.  If this
force is large enough (i.e. if the concentration of elastic energy
$f$ in the rod is larger then an activation threshold), the creep
starts - inelastic deformation that goes on for some time until the
rod brakes. Thus, at the moment when the inelastic strain $\eta(t)$
starts growing from zero, there should be a supply of strain energy
obtained from the work of the stress $\sigma_{zz}=\frac{\mathcal
F}{A(t)}$ on {\bf elastic deformation}. Denote by $f_{in}$ this
strain energy (of initiation).\par

 During the creep the homogeneous component $\sigma_{zz}$ of stress is equal
\begin{equation}
\sigma_{zz}(t)=\frac{{\mathcal F}}{A(t)}=\frac{{\mathcal
F}}{\lambda_{v}(t)^{2}\lambda_{d}^{-1}(t)A_{0}}=
\frac{\lambda_{d}(t){\mathcal F}}{A_{0}}= \frac{e^{\eta }{\mathcal
F}}{A_{0}}.
\end{equation}
The last equality is true given the assumption (natural for a
conventional creep) that {\bf inelastic volume change is
negligible}, i.e. $\lambda_{v}=1,\ \xi =0$., for a constant force
$\mathcal F$ and variable cross-section area $A(t)$.\par

Using the Hooke's law one can show that
$\epsilon_{z}=\frac{\sigma_{zz}}{Y} $ and that the strain energy is
equal to $f=\frac{{\mathcal F}^{2}e^{2\eta}}{2YA(0)^2}$. \par

Calculating the work of the load $\mathcal F$ on the total way
$L(e^{\eta (t)}-1)$ of the right end of the rod we get the
additional term in the Lagrangian (work of load on the inelastic
deformation) equal $\frac{\mathcal F}{A_{0}}(e^{\eta}-1)$.
\par
Overall action takes the form ($\xi(t)=0$)
 \beq
{\mathcal L}(S(t),\eta(t)) =[F_{CR}(S,\eta )+\frac{\mathcal
F}{A_{0}}(e^\eta -1)+\phi(S^{-1} \eta_{t}) +\Lambda {\mathcal
F}^{2}e^{2\eta }]Sdt,
\eeq
where $F(S,\eta)$ is the same as for
stress relaxation.\par

Acting as before we get the following dynamical system for
parameters $\eta ,S$:
\begin{equation}
\begin{cases}
\eta_{t}&= S\psi^{-1}(\frac{\mathcal F}{A_{0}}(e^\eta -1)+\Lambda
{\mathcal F}^{2}e^{2\eta }+
(q_{1}+2q_{2}S+b_{1}\eta +a_{1}\eta^2)),\\
S_{t}&=\frac{b_{0}}{2q_{2}}\psi^{-1}(\frac{\mathcal F}{A_{0}}(e^\eta
-1)+\Lambda {\mathcal F}^{2}e^{2\eta }+(q_{1}+2q_{2}S+b_{1}\eta
+a_{1}\eta^2)).
\end{cases}
\end{equation}
where $\psi(x)=(e^{Dx}-1)_{+}$ function as in the case of stress
relaxation.  We have in this system $\eta_{t}\geqq 0, S_{t}\geqq 0$
for the admissible initial values $S(0),\eta(0)$.
\par
 For the creep to start, the argument
of the function $\psi (x)$ in the system should be positive at
initial moment.
 Since $\sigma_{zz}=\frac{{\mathcal
F}e^{\eta}}{A_{0}}$, this condition takes the form
\begin{equation}
\frac{\sigma(0)^2}{2Y }+q_{1}+2q_{2}S(0)+a_{1}\eta(0)^{2}>0.
\end{equation}
Since $q_{1},q_{2},a_{1}$ are negative parameters, the inequality
(11.17) defines the stress (or strain energy - $f_{in}$) threshold
for the initiation of the creep processes (see \cite{HK}, p.7).
\begin{figure}[h]
\scalebox{.50}{\includegraphics{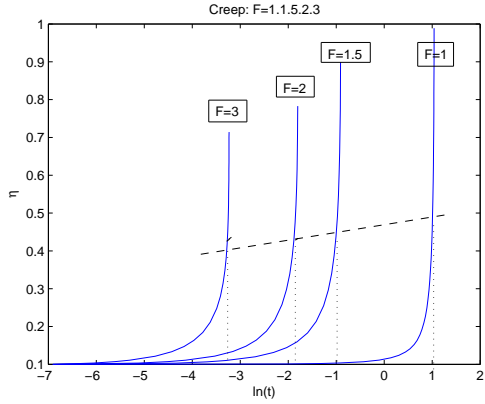}} \caption{Creep:
inelastic strain $\eta(t)$ for $F=1,1.5,2$ and 3.}
\end{figure}
  On the Figure 11 the graphs of $\eta(t)$ for the creep
are presented for different values of the force
$\frac{F}{A_{0}}=1,1.5,2$ ($A_{0}=1$).  There is a point on each
trajectory, corresponding to the instability of creep deformation
where the cross-section of the rod diminishes to zero and the rod
fails in so called ductile manner. As it can be seen from the
results of calculations, the higher is the force, the faster creep
deformation develops and the time to the ductile failure becomes
significantly shorter, for example, three times increase in force
results in more then 3 order of magnitude in time to failure.\par
 Comparing these graphs with the
experimental data (\cite{HK,BS}) we see the good qualitative (and,
for some materials, quantitative) agreement.

\section{Appendix A. Strain energy as a perturbation of the "ground state energy".}

In this section we discuss perturbation scheme of a ground state
Lagrangian $F(E^{in},S,{\bar N}),\
E^{in}=\frac{1}{2}\ln(g_{0}^{-1}g)$ by elastic deformation,
assuming that elastic strain tensor is small in compare to the
inelastic one.\par
 In the pure inelastic mode of behavior (free aging, special load
that produces $C(\phi )=g$) only metric quantities $g,S,{\bar N}$
enter the total Lagrangian $L$. Under the general load the material
metric $g$ is deformed into $C(\phi )$ and total deformation
$E^{tot}=\frac{1}{2}\ln(g_{0}^{-1}C(\phi))$ takes the place of
$E^{in}$.\par
 Assuming that $g^{-1}C(\phi )\approx I+small$, so that
$E^{el} \ll E^{in}$ we decompose $g^{-1}_{0}C(\phi
)=(g^{-1}_{0}g)\cdot (g^{-1}C(\phi)).$ Taking logarithm, we obtain

\[
E^{tot}=\frac{1}{2}\ln(exp(2E^{in})\cdot exp(2E^{el}))\simeq
\frac{1}{2}\ln(exp(2E^{in})\cdot (I+2E^{el}))
\]
Here we've used the linear approximation $exp(2E^{el})\simeq
I+2E^{el}$. Using the Campbell- Hausdorff-Dynkin formula we get
\begin{equation}
E^{tot}\simeq
\frac{1}{2}[\ln(exp(2E^{in}))+Ad(exp(2E^{in}))2E^{el}]=E^{in}+Ad(exp(2E^{in}))E^{el}.
\end{equation}
\par
We consider the "total ground state" Lagrangian $F$ depending on
 $E^{tot}$  through its invariants $I_{k}(E^{tot}),\ k=1,2,3$. Then we
 decompose it into Taylor
series by considering $E^{el}$ small in compare to $E^{in}$.  The
approximate expression for $E^{tot}$ above gives us up to the second
order terms
\begin{equation}
I_{k}(E^{tot})\simeq I_{k}(E^{in})+dI_{k}(E^{in})(E^{el\
g_{0}^{-1}g})+ d^{2}I_{k}(E^{in})(E^{el\ g_{0}^{-1}g},E^{el\
g_{0}^{-1}g})+h.o.t..
\end{equation}
Substituting this into the metric term $F(I_{k}(E^{tot})$ (for
this discussion we suppress in $F$ all other arguments it depends
on) we get, after recombining its terms, its decomposition up to
the second order
\begin{multline}
F(E^{tot})\simeq F(E^{in})+\sum_{k}F_{,I_{k}}(E^{in})<dI_{k}(E^{in}),E^{el\ g_{0}^{-1}g}>+\\
+\{ \sum_{k}F_{,I_{k}}(E^{in})<d^{2}I_{k}(E^{in})(E^{el\
g_{0}^{-1}g},E^{el\ g_{0}^{-1}g})+\\
+\sum_{ij}F_{I_{i}I_{j}}(E^{in}) <dI_{i}(E^{in}),E^{el\
g_{0}^{-1}g}><dI_{j}(E^{in}),E^{el\ g_{0}^{-1}g}> \}
\end{multline}
\par

 In this formula first term on the right is the basic metric
 ("ground state") energy describing, in particular, equilibrium values for the
metric $g$ (see example below).\par
 The second
term, linear by $E^{el}$ describes the interaction of elastic and
inelastic processes. In the absence of such interactions or other
material processes $g$ takes the value delivering minimum to the
basic energy $F(E^{in})$. Therefore, its differential takes a zero
value for corresponding value of the argument $E^{in}$ and linear
term vanishes.\par

Finally, the quadratic form in formula (12.3) is the conventional
elastic (strain) energy with variable and possible inhomogeneous
elasticity tensor.\par
 During the active processes of
configurational changes in the material (aging, in the zone of phase
transition) this elasticity tensor as well as the basic energy plays
an active role in the evolution. But when such processes stops (no
aging happens or wave of phase transition passed) and the material
metric $g$ is locked in some stable state (local minimum of $F$?),
the value of this tensor is also locked at the corresponding value
(see below).\par

In order to calculate the elasticity tensor $e$ we have to
calculate differentials of invariants $I_{k}$ of the inelastic
strain tensor $E^{in}$. We choose momenta $Tr(E^{k})$ as the basic
invariants of a (1,1)-tensors (\cite{TN}). In our, 3D case we have
$I_{1}(E)=Tr(E), I_{2}(E)=Tr(E^{2}), I_{3}(E)=det(E).$ Thus, we
have for its first differentials (\cite{TN})
\[
\frac{\partial I_{1}(E)}{E^{I}_{J}}=\delta ^{J}_{I};\
\frac{\partial I_{2}(E)}{E^{I}_{J}}=2E^{J}_{I};\ \frac{\partial
I_{3}(E)}{E^{I}_{J}}=3E^{J}_{K}E^{K}_{I}=3E^{2\ J}_{I}.
\]
and $ <dI_{1}(E),P>=Tr(P);\ <dI_{2}(E),P>=2Tr(EP);\
<dI_{3}(E),P>=3Tr(E^{2}P).$  Here we are using multiplication of
(1,1)-tensors. The second differentials of momenta have the form $
d^{2}I_{1}(E)=0;\ d^{2}I_{2}(E)(B,B)=2Tr(B^{2});\
d^{2}I_{3}(E)(B,B)=6Tr(EB^{2}). $
\par
Before using these expressions for differentials we notice that
since $g_{0}^{-1}g=exp(2E^{in})$, $(E^{in})^{g_{0}^{-1}g}=E^{in}$
and due to the properties of $Tr$ for all natural powers $a,b$ one
has $ Tr(E^{in\ a}((E^{el})^{g_{0}^{-1}g\ })^{b})=Tr(E^{in \ a}E^{el
\ b}). $ Thus, conjugation by $g_{0}^{-1}g$ disappear from the
formulas for Elastic energy.\par

Substituting expression for differentials in (12.3) we get
expression for $F(E^{tot})$ as the sum of "constant", linear by $\el
$ and quadratic by $\el $ terms
\begin{equation}
F(E^{tot})\simeq F(E^{in})+Tr(C \el )+Tr({\mathbf e}:\el :\el)
\end{equation}
Here $C$ is the (1,1)-tensor
\begin{equation}
C=F_{,I_{1}}(E^{in})I +2F_{,I_{2}}(E^{in})\inel
+3F_{,I_{3}}(E^{in})(\inel )^{2},
\end{equation}
and $\mathbf e$ the elasticity tensor
\begin{multline}
{\mathbf e}^{BD}_{AC}=2F_{,I_{2}}(E^{in})\delta
^{B}_{C}\delta^{D}_{A}+ 6F_{,I_{3}}(\inel )^{B}_{C}\delta^{D}_{A}+
 2F_{,I_{1}I_{1}}(\inel ) \delta ^{B}_{A}\delta^{D}_{C}+
 4F_{,I_{2}I_{2}}(\inel )^{B}_{A}(\inel )^{D}_{C}+\\
9F_{,I_{3}I_{3}}(\inel )^{2\ B}_{A}(\inel )^{2\ D}_{C}+
4F_{,I_{1}I_{2}}\delta^{B}_{A}(\inel )^{D}_{C}+
6F_{,I_{1}I_{3}}\delta^{B}_{A}(\inel )^{2\ D}_{C}+
12F_{,I_{2}I_{3}}(\inel )^{B}_{A}(\inel )^{2\ D}_{C}.
\end{multline}
If the ground state function $F$ is given as the function of
variables $g,S,{\bar N}$, these formulas determine values of elastic
moduli in a material which depend on the point $X$ and on time $t$
through the metric variables $g,S,{\bar N}$. If these variables take
stationary values,we get isotropic but nonhomogeneous material. If
they are constant - we return to the conventional linear elasticity.
\begin{example} {\bf Isotropic material} For isotropic material of linear
elasticity, elastic tensor (in its (1,1)-version) has the form
(\cite{MH})
\begin{equation}
e^{jl}_{ik}=2\mu \delta ^{l}_{i}\delta ^{j}_{k}+\lambda \delta
^{j}_{i}\delta ^{l}_{k}.
\end{equation}
Comparing with (15.7) we see immediately that there are two simple
cases when (15.3) determines an isotropic material.\par

{\bf Case 1 - generic.} Take $E^{in\ A}_{B}=h\delta ^{A}_{B}$ where
$h$ is a scalar function of $g,S,{\bar N}$. In this case
\begin{multline}
e^{BD}_{AC}=[2F_{,I_{2}}+6F_{,I_{3}}h]\delta ^{B}_{C}\delta ^{D}_{A}+\\
+[2F_{,I_{1}I_{1}}+4F_{,I_{1}I_{2}}h^{2}+9F_{,I_{3}I_{3}}h^{4}+4F_{,I_{1}I_{2}}h+6F_{,I_{1}I_{3}}h^{2}
+12F_{,I_{2}I_{3}}h^{3} ]\delta _{A}^{B}\delta ^{D}_{C}.
\end{multline}
The first bracket gives expression for $2\mu $ while the second one
- for $\lambda $.\par

{\bf Case 2 - simple elasticity}. In this case we have no aging,
$E^{in}=0$. Then we get material with
\begin{equation}
e^{BD}_{AC}=2F_{,I_{2}}(0)\delta ^{B}_{C}\delta
^{D}_{A}+2F_{I_{1}I_{2}}(0)\delta ^{B}_{A}\delta ^{D}_{C}.
\end{equation}
Thus, in this, restricted case $2\mu =2F_{,I_{2}}(0)$,$\lambda
=2F_{I_{1}I_{2}}(0).$
\end{example}
\begin{example}
Consider the model 1D case with one component of strain tensors
$\el , \inel $, trivial decomposition $E^{tot}=\el +\inel $ and
simple Taylor decomposition of the basic energy function
$F(E^{tot})$:
\begin{equation}
F(E^{tot}=F(\inel )+F'(\inel )\el +\frac{1}{2}F''(\inel )(\el
)^{2}+h.o.t.
\end{equation}
As a result, strain energy here has the form
\begin{equation}
U(\el )=\frac{1}{2}F''(\inel )(\el )^{2},
\end{equation}
and the Young's module (or compressional stiffness, in a case of an
elastic bar) is
\[
Y=\frac{1}{2}F''(\inel ).
\]
Consider two special cases.\par

1. Classical elasticity.  In this case we take
$F(E)=F_{0}+cE^{2}.$ In this case there is one equilibrium -
minimum $E=0$ that corresponds, for $E=\inel
=\frac{1}{2}ln(g_{0}^{-1}g)$ to the value $g(X,T)=g_{0}$
-constant.  We have $Y=c.$\par

2. Two-phase material (material that can exist in two stable
phases). In this example function $F(E)$ has two (locally) stable
states $g=g_{0},g_{1}$, or $E=0,\
E=Q=\frac{1}{2}\ln(g_{0}^{-1}g_{1})$
\begin{equation}
F(E)=F_{0}+\frac{1}{4}c_{4}E^{2}(E+Q^{-1})^{2}+\frac{1}{2}c_{2}E^{2}.
\end{equation}
Then, for $E=E^{in} $,
\begin{equation}
F(E)=F_{0}+\frac{1}{4}c_{4}(\frac{1}{2}\ln(g_{0}^{-1}g) )
^{2}(\frac{1}{2}\ln(g_{1}^{-1}g))^{2}+\frac{1}{2}c_{2}(\frac{1}{2}\ln(g_{0}^{-1}g))^{2}.
\end{equation}
We have
\[
F''(E)=c_{2}+\frac{1}{2}c_{4}((E+Q^{-1})^{2}+4E(E+Q^{-1})+E^{2}).
\]
Young module in the state $g=g_{0}$ is equal to
\[ Y_{0}=c_{2}+\frac{1}{2}c_{4}Q^{-2},
\]
while in the second stable state $g=g_{1}$,
\[
Y_{1}=c_{2}+\frac{1}{2}c_{4}((Q+Q^{-1})^{2}+4Q(Q+Q^{-1})+Q^{2})=Y_{0}+3c_{4}(1+Q^{2})
\]
Thus, in a case of a wave of phase transition going along the bar,
Young module changes by the amount $ Y_{1}-Y_{0}=3c_{4}(1+Q^{2}). $
\end{example}

\section{Appendix II. Variations}
Variations of some expressions for the Lagrangian (5.4-5.5) are
calculated here and presented in a table.  All terms in Lagrangian
density $L(G,\phi)$ will be refereed to the mass form
$dM=\rho_{0}d_{G}V= \rho_{0}S\sqrt{\vert g\vert}d^{4}X$. In other
terms we calculate variation $\int f(A)dM$ by $A$. Result of
variation has the form ${\mathcal V}dM$ : $ \delta f(A)dM =
{\mathcal V}dM.$ In the calculations we repeatedly using the
following standard relation $\delta \sqrt{\g }=\frac{\sqrt{\g
}}{2}g^{IJ}\delta g_{IJ}$ (see, for instance, \cite{ADM}).
 In the table below we present tensors $\mathcal V$ for different
 $f$.\par

As an example of such a calculation we provide calculation of
$\delta\  [div_{g}({\vec N})dM]$ where
$div_{g}(N)=\frac{1}{\sqrt{\vert g\vert}}\frac{\partial}{\partial
X^I}(\sqrt{\vert g\vert}N^I )$:

\begin{multline}
\delta div_{g}({\vec N})dM=div_{g}({\vec N})\rho_{0}( \sqrt{\vert
g\vert }\delta S+S\frac{\sqrt{\vert g\vert }}{2}g^{IJ}\delta
g_{IJ})+\delta (\frac{1}{\sqrt{\vert
g\vert}}\frac{\partial}{\partial
X^I}(\sqrt{\vert g\vert}N^I ))dM=\\
=(div_{g}({\vec N})S^{-1}\delta S- \frac{\partial }{\partial
X^I}ln(\rho_{0}S)\delta N^{I}+ \left[div_{g}({\vec N})
\frac{1}{2}-\frac{1}{2}(div_{g}({\vec
N})-\frac{1}{2}N^{I}\frac{\partial
}{\partial X^I}ln(\rho_{0}S)) \right] g^{AB}\delta g_{AB})dM=\\
(div_{g}({\vec N})S^{-1}\delta S- \frac{\partial }{\partial
X^I}ln(\rho_{0}S)\delta N^{I}+
\left[-\frac{1}{2}N^{I}\frac{\partial }{\partial
X^I}ln(\rho_{0}S)) \right] g^{AB}\delta g_{AB})dM
\end{multline}
since
\begin{multline}
\delta (\frac{1}{\sqrt{\vert g\vert}}\frac{\partial}{\partial
X^I}(\sqrt{\vert g\vert}N^I ))dM=-(\frac{1}{\sqrt{\vert
g\vert}^2}\delta \sqrt{\vert g\vert }\frac{\partial}{\partial
X^I}(\sqrt{\vert g\vert}N^I ))dM-\frac{\partial }{\partial
X^I}(\rho_{0}S)\delta(\sqrt{\vert g\vert}N^I )d^{4}X=\\
=[-\frac{1}{2}(div_{g}({\vec N})-\frac{1}{2}N^{I}\frac{\partial
}{\partial X^I}ln(\rho_{0}S))g^{AB}\delta g_{AB}-\frac{\partial
}{\partial X^I}ln(\rho_{0}S)\delta N^{I}]dM
\end{multline}
\par
Formula of variation of $hR(g)\sqrt{\g}$ in the 5th row of the table
is taken from \cite{FM}, Prop.3.2.\par

To find variation of the strain energy density
$f(G,E^{in})\rho_{0}S\sqrt{\g }d^{4}X$ we first take variation of
$dM=\rho_{0}S\sqrt{\g }d^{4}X$ to get the first two terms in the
last row of the Table, then - explicit variation by $g$ if the
strain energy function $f$ depends on $g$ not just through
$E^{el}$.  Finally for variation by $g$ through the strain tensor
${\el}^{I}_{J}=\frac{1}{2}g^{IK}(C(\phi )_{KJ}-g_{KJ})$ we have

\begin{multline} \delta f(\el ) =\frac{\partial f}{\partial
{\el}^{I}_{J}}\delta {\el}^{I}_{J}=\frac{1}{2}\frac{\partial
f}{\partial {\el}^{I}_{J}}\delta g^{IK} C(\phi
)_{KJ}=\frac{1}{2}\frac{\partial f}{\partial
{\el}^{I}_{J}}[-g^{IA}g^{KB}\delta g_{AB}] C(\phi )_{KJ}=\\
-\frac{1}{4}\frac{\partial f}{\partial
{\el}^{I}_{J}}[(g^{IA}g^{KB}+g^{IB}g^{KA})\delta g_{AB}] C(\phi
)_{KJ}\delta g_{AB}
 =-\frac{1}{4}\frac{\partial f}{\partial
{\el}_{MN}}[\left( \delta^{J}_{M}g_{IN}+\delta^{J}_{N}g_{IM} \right)
\times \\ \times \left((g^{IA}g^{KB}+g^{IB}g^{KA})\delta
g_{AB}\right)] C(\phi )_{KJ}\delta g_{AB}= -\frac{1}{4}(
\frac{\partial f}{\partial {\el}_{JA}}g^{KB}+ \frac{\partial
f}{\partial {\el}_{JB}}g^{KA}+ \\ +\frac{\partial f}{\partial
{\el}_{AJ}}g^{KB}+\frac{\partial f}{\partial {\el}_{BJ}}g^{KA})
C(\phi )_{KJ}\delta g_{AB}= -\frac{1}{2}S^{(AB)}\delta g_{AB},
 \end{multline}
 where $S^{(AB)}$ is the symmetrization of the second Piola-Kirchoff
 Tensor $S$ (see \cite{MH}), the last equality is proved in \cite{CP1}.
\begin{table}
\begin{center}
\begin{tabular}{| l | c | }
\hline Term & Variation by $S,\ N^I,\ g_{IJ}$\\ \hline

$f(\g )$ & $S^{-1}f(\g )\delta S+ [\frac{1}{2}f(\g )g^{IJ}+f'(\g )
{\vert g\vert}g^{IJ}]\delta g_{IJ}$  \\
\hline

$f(S)$   & $[f'(S)+S^{-1}f(S)]\delta S+\frac{1}{2}f(S)g^{IJ}\delta g_{IJ}$  \\
\hline

$\Vert {\vec N} \Vert^{2}_{g}$ & $\Vert {\vec N}
\Vert^{2}_{g}S^{-1}\delta S+2N_{I}\delta N^I+ [N^{I}N^{J}+
\frac{1}{2}\Vert {\vec N} \Vert^{2}_{g}g^{IJ}]\delta g_{IJ}$   \\
\hline
 $\chi (K^{I}_{J})$ &
\parbox[c]{12cm}{ {\small
$S^{-1}K^{I}_{J}\frac{\partial \chi(K)}{\partial K^{I}_{J}}\delta S+
[ -S^{-1}\frac{\partial \chi }{\partial
K^{A}_{J}}g^{AS}\partial_{X^{I}}g_{SJ}+ \frac{1}{\rho_{0}S\sqrt{\g
}} \partial_{X^{S}}(\rho_{0}\sqrt{\g }
 \frac{\partial \chi }{\partial K^{A}_{J}}g^{AS}g_{IJ})+$}\\
{\small $+\frac{1}{\rho_{0}S\sqrt{\vert g\vert
}}\partial_{X^{J}}(\rho_{0}\sqrt{\g} \frac{\partial \chi }{\partial
K^{I}_{J}})]\delta N^I+[-\frac{\partial \chi }{\partial K^{I}_{J}}
S^{-1}g^{IA}\frac{\partial N^{B}}{\partial X^{J}}-\frac{\partial
\chi }{\partial K^{I}_{B}} S^{-1}g^{IS}\frac{\partial
N^{A}}{\partial X^{S}}- \frac{\partial \chi }{\partial
K^{I}_{J}}g^{IA}K^{B}_{J}- \frac{1}{\rho_{0}S\sqrt{\g
}}\partial_{t}\left( \rho_{0}\g \frac{\partial \chi }{\partial
K^{I}_{B}}g^{IA}\right) +
 \frac{1}{\rho_{0}S\sqrt{\g}}\partial_{X^{K}} \partial_{t}\left(
\rho_{0}\g \frac{\partial \chi }{\partial K^{I}_{B}}g^{IA} N^{K}
\right) ]\delta g_{AB}$}} \\
\hline

$h R(g)$ & $S^{-1}h R(g)\delta S+ [-h(Ric(g)-\frac{1}{2}R(g)g)+
\frac{1}{\rho_{0} S}[\Delta_{g}(\rho_{0}h S)g^{AB}+
Hess(\rho_{0}h S) ]]\delta g_{AB}$   \\
\hline

$div_{g}(\vec N)$ & $div_{g}({\vec N})S^{-1}\delta S- \frac{\partial
}{\partial X^K}ln(\rho_{0}S)\delta N^{K}+
\left[-\frac{1}{2}N^{K}\frac{\partial }{\partial
X^K}ln(\rho_{0}S)) \right] g^{IJ}\delta g_{IJ}$ \\
\hline

$f(G,E^{el})$ & $S^{-1}f\delta S+  [\frac{1}{2}f g^{AB}+
\frac{\partial f}{\partial g_{AB}}\text{\Small exp}-\frac{1}{2}S^{(AB)}]\delta g_{AB}$\\
\hline
\end{tabular}
\caption{Table of Variations.}\label{Ta:first}
\end{center}
\end{table}

\section{Conclusion}
In this work, we consider the intrinsic material metric tensor to be
an additional parameter of state, i.e., an internal variable that
characterizes material degradation and aging. The material metric
tensor $G$ is a conjugate (with respect to a particular Lagrangian)
to the canonical Energy-Momentum Tensor (or to the Eshelby
energy-stress tensor to some degree).\par

Equations of metric evolution, (i.e., the aging equations), are
derived as the Euler-Lagrange equation of a corresponding
variational problem.  Canonical energy-momentum tensor (or Eshelby
Tensor) play a role of the source of metric evolution. This
represents an alternative approach to numerous phenomenological
damage models, which usually have more adjustable parameters than
practical testing is able to determine. Thus it is difficult to
validate the models since they can almost always be adjusted to
reach an agreement with the experiment. In contrast, a variational
approach prescribes a functional form of the aging equations, limits
the number of constants (adjustable parameters) employed in the
Lagrangian, provides a simple physical interpretation of the
constants, and admits an essential experimental examination of the
validity of the basic assumptions of the model. Particular examples
(aging homogeneous rod, see Sec. 11 or [20], cold drawing (necking)
[18], residual stress and others) can be analyzed theoretically and
unambiguously tested in the experiments as a natural continuation of
the present work.

A natural development of this scheme requires the following:
thermodynamical interpretation of the balance equation considered in
section 10, especially the structural entropy evolution manifested
by the increase of the lapse function $S(t)$ during aging;
introduction of a temperature dependence of the material metric
(based on an unpublished work by A.Chudnovsky and B.Kunin);
development of models ("ground energy" $F$ + kinetic potential $\hat
A$ + possibly other metrical terms) characterizing a hierarchy of
aging phenomena for specific materials; and especially, the
development of models of phase transition (front propagation,
fractal restructuring of materials, etc.).

We would like to express our gratitude to Professor M. Francaviglia
for his attention to this work. We would also like to thank
Professor R.Tucker and the participants of his seminar in the
Physics Department of Lancaster University,UK for the useful
discussion.

\end{document}